\documentclass[fleqn,usenatbib]{mnras}

\usepackage{newtxtext,newtxmath}

\usepackage[T1]{fontenc}

\DeclareRobustCommand{\VAN}[3]{#2}
\let\VANthebibliography\thebibliography
\def\thebibliography{\DeclareRobustCommand{\VAN}[3]{##3}\VANthebibliography}

\usepackage{graphicx}	
\usepackage{amsmath}	

\title[MCAO residuals and astrometry]{Temporal spectrum of multi-conjugate adaptive optics residuals and impact of tip-tilt anisoplanatism on astrometric observations}

\author[G. Carlà et al.]{
Giulia Carlà,$^{1}$\thanks{E-mail: giulia.carla@inaf.it}
Cédric Plantet,$^{1}$
Lorenzo Busoni$^{1}$
Guido Agapito$^{1}$
\\
$^{1}$INAF, Osservatorio Astrofisico di Arcetri, Largo Enrico Fermi 5, 50125 Firenze, Italy
}

\date{Accepted XXX. Received YYY; in original form ZZZ}

\pubyear{2022}

\begin{document}
\label{firstpage}
\pagerange{\pageref{firstpage}--\pageref{lastpage}}
\maketitle

\begin{abstract}
Multi-conjugate adaptive optics (MCAO) will assist a new era of ground-based astronomical observations with the extremely large telescopes and the Very Large Telescope. High precision relative astrometry is among the main science drivers of these systems and challenging requirements have been set for the astrometric measurements. A clear understanding of the astrometric error budget is needed and the impact of the MCAO correction has to be taken into account. In this context, we propose an analytical formulation to estimate the residual phase produced by an MCAO correction in any direction of the scientific field of view. The residual phase, computed in the temporal frequency domain, allows to consider the temporal filtering of the turbulent phase from the MCAO loop and to extract the temporal spectrum of the residuals, as well as to include other temporal effects such as the scientific integration time. The formulation is kept general and allows to consider specific frameworks by setting the telescope diameter, the turbulence profile, the guide stars constellation, the deformable mirrors configuration, the modes sensed and corrected and the tomographic reconstruction algorithm. The formalism is presented for both a closed loop and a pseudo-open loop control. We use our results to investigate the effect of tip-tilt residuals on MCAO-assisted astrometric observations. We derive an expression for the differential tilt jitter power spectrum that also includes the dependence on the scientific exposure time. Finally, we investigate the contribution of the differential tilt jitter error on the future astrometric observations with MAVIS and MAORY.
\end{abstract}




\section{Introduction}
The next generation of ground-based telescopes equipped with adaptive optics (AO) will provide unprecedented resolutions to astronomical observations in the visible and near infrared wavelengths. This is the case of the extremely large telescopes, the new class of 25-40m telescopes observing in the near infrared \citep{elt,gmt,tmt}, as well as the 8m Very Large Telescope (VLT) observing in the visible \citep{vltaof}. Most of the mentioned telescopes foresee the use of multi-conjugate adaptive optics (MCAO) \citep{Beckers88,Rigaut18} modules to compensate for the wavefront distortions induced by atmospheric turbulence: MAORY \citep{maory} for the Extremely Large Telescope (ELT), NFIRAOS \citep{nfiraos} for the Thirty Meter Telescope and MAVIS \citep{Rigaut20} for the VLT. This flavour of adaptive optics aims to overcome the anisoplanatism problem, that represents a major limitation for single-conjugated adaptive optics (SCAO) \citep{Chassat89,Fried82}, through the use of both multiple guide stars (GSs) and deformable mirrors (DMs). The tomographic reconstruction of the turbulent volume from the GSs and the compensation for different layers of the atmosphere by the DMs help increase the isoplanatic patch, allowing the MCAO correction to provide uniform diffraction limited images over wide fields of view. The high angular resolution, the uniformity of the correction over wide areas, the large number of reference sources with high image quality provided and the control of the field distortions through the DMs conjugated in altitude are characteristics that make MCAO a good candidate for astrometric observations. High precision relative astrometry is, indeed, one of the main science drivers of the instruments equipped by the mentioned MCAO modules. The limiting astrometric precision is given by the centroiding error \citep{Lindegren78} and leads to the challenging requirements that have been set for these systems: 50µas of astrometric precision for MAORY (goal of 10µas, \citealt{Rodeghiero19}), 150µas for MAVIS (goal of 50µas, \citealt{Monty21}) and 50µas for NFIRAOS (goal of 10µas, \citealt{nfiraos_astrometry}). It is then crucial to investigate all possible sources of error in order to keep the astrometric error budget within this fundamental limitation. An exhaustive list of the main contributions to the astrometric error in the case of MCAO-assisted observations was provided in \citet{Trippe10}. Among the sources of error mentioned, we are interested in investigating tip-tilt atmospheric residuals. In general, tip-tilt residuals affect the astrometric precision by introducing fluctuations of the position of a source with respect to the nominal position on the detector. On the one hand, the amount of fluctuations integrated during the individual exposure can determine an increasing of the size and a change in shape of the point spread function (PSF), with typical PSF elongation effect; on the other hand, if the fluctuations are not totally integrated within the exposure time of the image, a jitter of the source position can also be observed between successive frames. Relative astrometry, intended as the measurement of the distance between two distinct sources, can be affected by both effects: the former contributes to the centroiding error in measuring the position of each object, while the latter leads to the \textit{differential tilt jitter} error, that is, the uncertainty in the distance measurement due to the relative residual jitter \citep{Fritz10,Cameron09}. The knowledge of the spatial and temporal dependence of tip-tilt residuals is needed to characterize the behavior of the related astrometric error. For SCAO systems, tip-tilt anisoplanatism is well known and has been thoroughly modeled: measuring tip-tilt through an off-axis reference determines a residual tip-tilt on the target that linearly increases with the separation between the two sources, the linear dependence on the distance being valid for each pair of objects in the field \citep{Sandler94,Sasiela94,Hardy98}. However, the characterization is more elaborate for the MCAO case, since the geometry with multiple guide stars and multiple DMs needs to be taken into account and can lead to complex behaviors. As pointed out in \citet{Trippe10}, tip-tilt anisoplanatism is not well understood for this flavour of adaptive optics and, to our knowledge, an analysis does not exist yet.
In this context, we propose an analytical formulation that allows the derivation of the temporal power spectral density (PSD) of the MCAO residual phase in any direction of the scientific field of view, by means of the spatio-temporal statistics of the turbulence-induced distortions and of the temporal transfer functions of an MCAO loop. The phase is intended as decomposed on a modal basis (e.g. Zernike modes, \citealt{Noll76}). Differently from existing approaches providing an estimation of MCAO residuals in the spatial frequency domain (e.g. \citealt{Neichel09}), the presented method evaluates, for each mode, the MCAO residual phase in the temporal frequency domain and allows to include temporal effects such as the scientific integration time. The formulas are general and allow to analyse specific frameworks depending on the telescope aperture, the turbulence profile, the natural guide star (NGS) or laser guide star (LGS) asterism, the number and conjugation heights of the DMs, the sensed and corrected modes of distortion. The control loop and the tomographic reconstruction algorithm can also be chosen: in particular, we provide expressions in the case of either a closed-loop or a pseudo-open loop control. 
We then specialize our results to NGS-based systems and we analyse the behavior of MCAO tip-tilt anisoplanatism. We model the effect on tip-tilt residuals of the scientific integration time as well. Moreover, we provide an analytical expression to derive the temporal PSD of differential tilt jitter. Finally, we show an application where we make use of the presented formulas to quantify the contribution of differential tilt jitter to the future MCAO-assisted astrometric observations, choosing MAORY and MAVIS as case studies.
\\
\noindent In Sec.~\ref{sec:performance}, we present the analytical approach and we derive the expression for the temporal PSD of the residual wavefront in the case of an MCAO correction; in Sec.~\ref{sec:aniso}, we use the formulas to analyse the spatial and temporal behavior of tip-tilt residuals, as well as to provide the expression for the differential tilt jitter error; in Sec.~\ref{sec:application}, we apply our results on differential tilt jitter to the MAORY and MAVIS cases.

\section{Temporal power spectral density of MCAO wavefront residuals}
\label{sec:performance}
The aim of this section is to derive an analytical expression of the residual phase produced by an MCAO correction in a generic direction of the field of view as a function of the temporal frequencies. From this quantity, the temporal power spectral density (PSD) of the residual phase can be derived as:
\begin{equation}
\label{eq:psd_res_1}
    S_{res}^{\alpha}(\nu) = \big \langle \phi_{res}^{\alpha}(\nu) \phi_{res}^{\alpha \: \dagger}(\nu) \big \rangle \, ,
\end{equation}
where $\alpha$ identifies the position in the field of view, $\nu$ is the temporal frequency, $\langle \cdot \rangle$ is the ensemble average, $^{\dagger}$ denotes the conjugate-transpose and $\phi$ represents the $\mathcal{L}$- or $Z$-transform of the phase, depending on whether a continuous or discrete-time domain is considered. From the integration of $S_{res}^{\alpha}$, the variance of the residual phase can be computed as well:
\begin{equation}
\label{eq:var_res_1}
    (\sigma_{res}^{\alpha})^2 = \int d\nu \: S_{res}^{\alpha}(\nu) \, .
\end{equation}
Among the sources of error contributing to the error budget of an MCAO correction, the presented method allows to take into account tomographic, noise and temporal errors.

\begin{figure}
    \centering
    \includegraphics[width=0.9\linewidth]{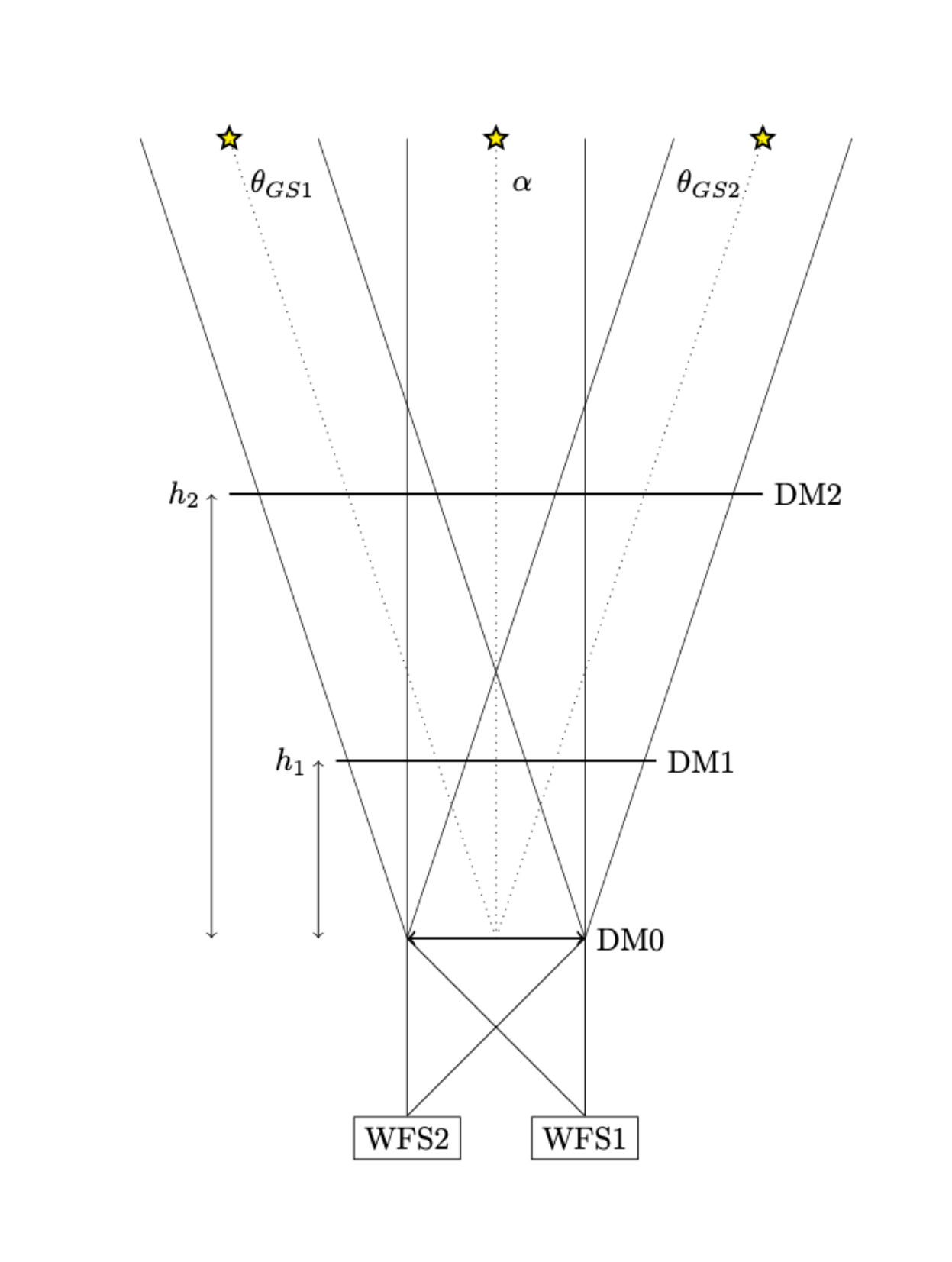}
    \caption{Scheme of the system geometry. In the example, there are two DMs conjugated at $h_1$ (DM1) and $h_2$ (DM2) and one at the ground layer (DM0), two guide stars at coordinates $\theta_{GS1}$ and $\theta_{GS2}$ and the scientific target at $\alpha$. The wavefront distortion is measured by WFS1 and WFS2 looking at, respectively, GS1 and GS2.}
    \label{fig:geometry}
\end{figure}

We consider the configuration in Fig.~\ref{fig:geometry}: the target and the guide stars (GSs) are, respectively, at positions $\alpha$ and $\bmath{\theta_{GS}} = [\theta_1, \theta_2, ..., \theta_{N_{GS}}]$ with respect to the telescope's axis.
The light from the sources passes through $N_l$ layers of atmospheric turbulence before arriving at the pupil of the telescope. The turbulent layers are assumed to follow Taylor's frozen flow hypothesis. The turbulence-induced distortions are considered as decomposed onto wavefront modes and are measured by $N_{GS}$ wavefront sensors (WFSs), each sensing $n$ modes, and corrected by $N_{DM}$ deformable mirrors optically conjugated at altitudes $h_{j=1}^{N_{DM}}$ and compensating a total of $m_{DM} = \sum^{N_{DM}}_{k=1} m_k$ modes. In the following, we will denote the turbulent and residual phase in the direction of the target as $\phi_{turb}^{\alpha}$ and $\phi_{res}^{\alpha}$ respectively, the turbulent and residual phase in the direction of the guide stars as $\phi_{turb}^{\bmath{\theta_{GS}}}$ and $\phi_{res}^{\bmath{\theta_{GS}}}$ respectively and the phase applied on the deformable mirrors as $\phi_{DM}$. It follows that $\phi_{turb}^{\alpha}$ and $\phi_{res}^{\alpha}$ are vectors of $n$ elements, $\phi_{turb}^{\bmath{\theta_{GS}}}$ and $\phi_{res}^{\bmath{\theta_{GS}}}$ are vectors of $(n\cdot N_{GS})$ elements and $\phi_{DM}$ is a vector of $m_{DM}$ elements.

\begin{figure}
    \centering
    \includegraphics[width=\linewidth]{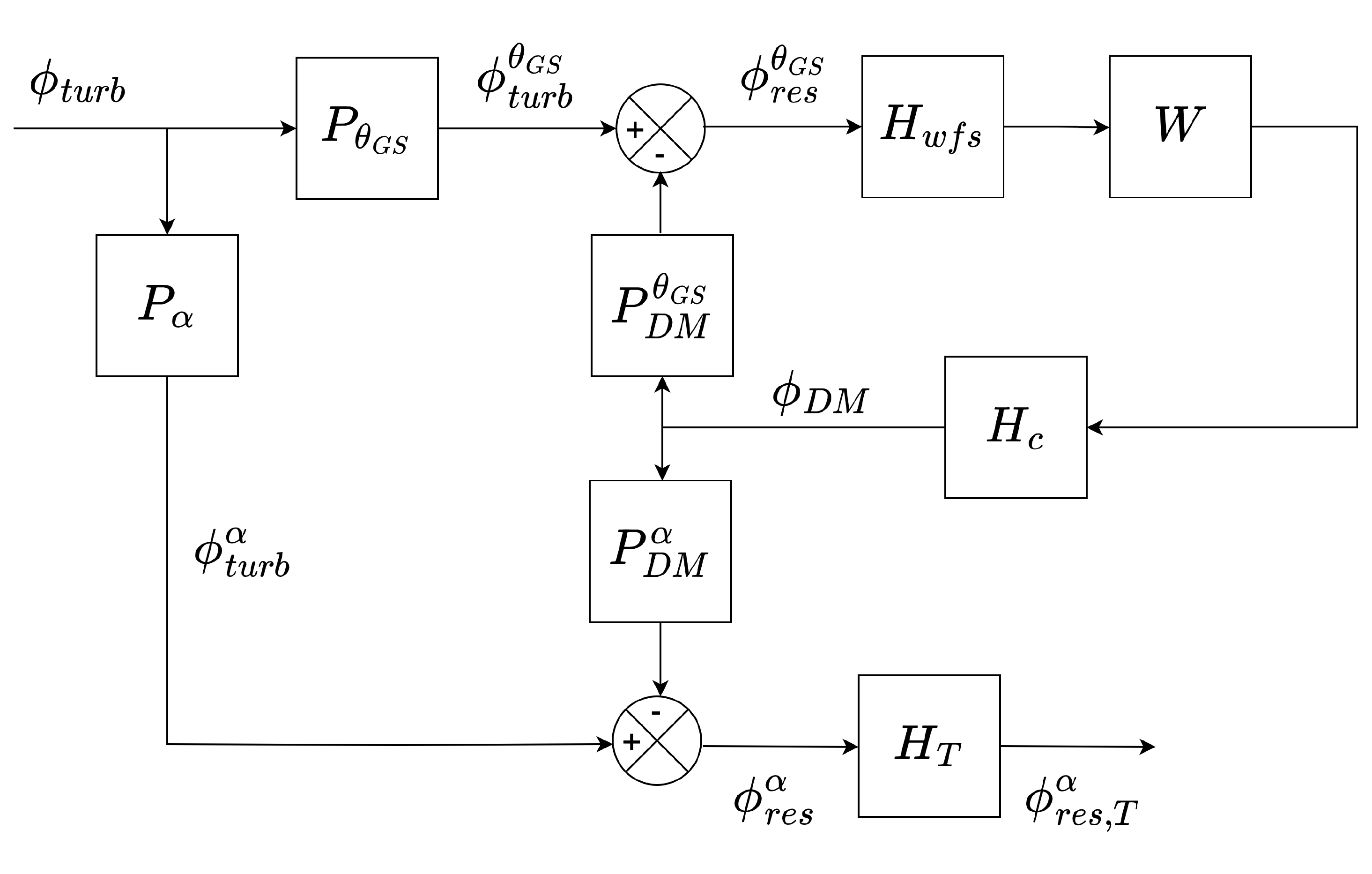}
    \caption{Diagram of the control loop. The phase on the DMs is controlled in closed loop from the measurements on the guide stars and its projection along $\alpha$ determines the residual phase on the target in $\alpha$. The $P_{\bmath{\theta_{GS}}}$ and $P_{\alpha}$ blocks have been introduced as projections of the turbulent phase onto $\bmath{\theta_{GS}}$ ($\phi_{turb}^{\bmath{\theta_{GS}}} = P_{\bmath{\theta_{GS}}} \phi_{turb}$) and $\alpha$ ($\phi_{turb}^{\alpha} = P_{\alpha} \phi_{turb}$) respectively. The $H_T$ block represents the temporal filtering by the scientific instrument, as it will be shown in Sec.~\ref{sec:sci_time}.} 
    \label{fig:loop_scheme}
\end{figure}

We start writing the residual phase along $\alpha$ as the difference between the turbulent phase and the correction phase, both evaluated in the direction of interest:
\begin{equation}
\begin{split}
\label{eq:phi_res_alpha_1}
\phi_{res}^{\alpha}(\nu) &= \phi_{turb}^{\alpha}(\nu) - \phi_{corr}^{\alpha}(\nu) \\
&= \phi_{turb}^{\alpha}(\nu) - P_{DM}^{\alpha} \phi_{DM}(\nu)\, ,
\end{split}
\end{equation}
where $\phi_{corr}^{\alpha}$ is the correction phase in the direction $\alpha$, obtained through the matrix $P_{DM}^{\alpha}$ of size $n\times m_{DM}$ that projects the modes on the DMs as seen in the direction $\alpha$ onto the pupil. In the SCAO case, $P_{DM}^{\alpha}$ is the identity for any direction $\alpha$ as the correction is common to all directions of the field of view ($\phi_{corr}^{\alpha} = \phi_{corr}$).\newline
We define $\phi_{DM}(\nu)$ as:
\begin{equation}
    \label{eq:phi_dm_1}
    \phi_{DM}(\nu) = H_{ol}(\nu) W \big(\phi_{res}^{\bmath{\theta_{GS}}}(\nu) + \phi_{n}(\nu) \big) \, ,
\end{equation}
where $H_{ol}$ is the open-loop transfer function of the AO feedback loop, $W$ is the reconstruction matrix, with dimension $m_{DM}\times (n\cdot N_{GS})$, relating the modes measured by the WFSs and the ones to be applied by the DMs and $\phi_{n}(\nu)$ is the WFSs measurement noise on the modes. We assumed ideal WFSs, meaning that they perform a direct measurement of the phase. In the case of a pure integrator, the expression of $H_{ol}$ is \citep{Madec99,Correia17}:
\begin{equation}
\begin{split}
    H_{ol}(s) &= H_{wfs}(s) H_c(s)\\
    &= \dfrac{(1 - e^{-sT})}{sT} \dfrac{g}{sT} e^{-s T_d} \, ,
\end{split}
\end{equation}
where we limited the contributors to the wavefront sensor and the control and where $s = i2\pi\nu$ is the Laplace variable, $g$ is the gain, $T = 1 / v_{loop}$ with $v_{loop}$ the loop frequency, $T_d$ is the delay time of the control and where we defined $H_{wfs}(s) = (1 - e^{-sT})/sT$ and $H_c(s) = g/sT e^{-s T_d}$.
\newline
By replacing Eq.~\eqref{eq:phi_dm_1} in Eq.~\eqref{eq:phi_res_alpha_1} as referred to the guide stars directions ($\alpha = \bmath{\theta_{GS}}$), we get an expression of the residual phase on the guide stars:
\begin{equation}
    \label{eq:phi_res_gs}
    \begin{split}
    \phi_{res}^{\bmath{\theta_{GS}}}(\nu) &= \big(Id + P_{DM}^{\bmath{\theta_{GS}}} H_{ol}(\nu) W \big)^{-1} \phi_{turb}^{\bmath{\theta_{GS}}}(\nu) \\
    & \:\:\:- \big(Id + P_{DM}^{\bmath{\theta_{GS}}} H_{ol}(\nu) W \big)^{-1} P_{DM}^{\bmath{\theta_{GS}}} H_{ol} (\nu) W \phi_{n}(\nu)  \\
    &= H_r(\nu)\phi_{turb}^{\bmath{\theta_{GS}}}(\nu) - H_n(\nu) \phi_{n}(\nu) \, ,
    \end{split}
\end{equation}
where $P_{DM}^{\bmath{\theta_{GS}}}$ is the DMs-WFSs projection matrix, with dimension $(n\cdot N_{GS})\times m_{DM}$ and $Id$ is an $(n\cdot N_{GS})\times (n\cdot N_{GS})$ identity matrix. We defined
\begin{equation}
\label{eq:rtf}
    H_r(\nu) = \big(Id + P_{DM}^{\bmath{\theta_{GS}}} H_{ol}(\nu) W \big)^{-1} \, 
\end{equation} 
as the Rejection Transfer Function (RTF), and
\begin{equation}
\label{eq:ntf}
H_n(\nu) = \big(Id + P_{DM}^{\bmath{\theta_{GS}}} H_{ol}(\nu) W \big)^{-1} P_{DM}^{\bmath{\theta_{GS}}} H_{ol}(\nu) W \, 
\end{equation}
as the Noise Transfer Function (NTF) of the MCAO loop. It is worth noting that these expressions also include a dependence on the spatial reconstruction. If taking the SCAO limit, $P_{DM}^{\bmath{\theta_{GS}}}$ and $W$ become equal to one and the classical definitions of RTF and NTF are retrieved \citep{Agapito17}. \newline
We then replace Eq.~\eqref{eq:phi_res_gs} in Eq.~\eqref{eq:phi_dm_1}:
\begin{equation}
\begin{split}
    \label{eq:phi_dm_2}
    \phi_{DM}(\nu) &= H_{ol}(\nu) W \big(H_r(\nu) \phi_{turb}^{\bmath{\theta_{GS}}}(\nu) - H_n(\nu) \phi_{n}(\nu) + \phi_{n}(\nu) \big) \\
    &= H_{ol}(\nu) W H_r(\nu) \big( \phi_{turb}^{\bmath{\theta_{GS}}}(\nu) + \phi_{n}(\nu) \big) \\
    &= H_{n,tomo}(\nu) \big(\phi_{turb}^{\bmath{\theta_{GS}}}(\nu) + \phi_{n}(\nu) \big) \, ,
\end{split}
\end{equation}
where we used the relation $H_r(\nu)+H_n(\nu) = Id$, as derived from the sum of Eq.~\eqref{eq:rtf} and Eq.~\eqref{eq:ntf}, and where we defined the matrix $H_{n,tomo}(\nu) = H_{ol}(\nu) W H_r(\nu)$ as a tomographic NTF.\newline
By substituting Eq.~\eqref{eq:phi_dm_2} in Eq.~\eqref{eq:phi_res_alpha_1}, we derive a final expression for the residual phase along $\alpha$:
\begin{equation}
    \label{eq:phi_res_alpha_2}
    \begin{split}
    \phi_{res}^{\alpha}(\nu) &= \phi_{turb}^{\alpha}(\nu) - P_{DM}^{\alpha}\Big[H_{n,tomo}(\nu)\big(\phi_{turb}^{\bmath{\theta_{GS}}}(\nu) + \phi_{n}(\nu) \big)\Big] \\
    &= \phi_{turb}^{\alpha}(\nu) - H_{n,tomo}^{\alpha}(\nu) \big(\phi_{turb}^{\bmath{\theta_{GS}}}(\nu) + \phi_{n}(\nu) \big) \, ,
    \end{split}
\end{equation}
where $H_{n,tomo}^{\alpha}(\nu)=P_{DM}^{\alpha} H_{n,tomo}(\nu)$ is the tomographic NTF projected along $\alpha$. The diagram of the control loop described is shown in Fig.~\ref{fig:loop_scheme}.

From Eq.~\eqref{eq:psd_res_1} and Eq.~\eqref{eq:phi_res_alpha_2} we can also compute the temporal power spectrum of the residual phase along $\alpha$:
\begin{equation}
    \begin{split}
    \label{eq:psd_res_2}
        S^{\alpha}_{res} (\nu) &= S_{turb}^{\alpha}(\nu) + H_{n,tomo}^{\alpha}(\nu) \big(S_{turb}^{\bmath{\theta_{GS}}}(\nu) + S_{n}(\nu) \big) H_{n,tomo}^{\alpha \: \dagger}(\nu) \\
        & \:\:\: - 2 Re \big( H_{n,tomo}^{\alpha}(\nu) S_{turb}^{\bmath{\theta_{GS}},\alpha}(\nu) \big) \, ,
    \end{split}
\end{equation}
where $S_{turb}^{\alpha}$ is the temporal PSD of the turbulence, $S_{turb}^{\bmath{\theta_{GS}}}$ is the temporal PSD of the turbulence on the guide stars directions, $S_{n}$ is the temporal PSD of the noise and $S_{turb}^{\bmath{\theta_{GS}},\alpha}$ is the Cross PSD (CPSD) \citep{Plantet22} of the turbulence between the guide stars and the target. We assumed turbulence and noise to be uncorrelated.\newline
The derived expression can provide a fast evaluation of the MCAO residuals in the field of view, given a statistics of turbulence and noise and the temporal filtering operated by the adaptive optics loop. It is worth noting that the SCAO limit of Eq.~\eqref{eq:psd_res_2} gives the same expression as provided in Eq.(54) of \citet{Plantet22}.

Another version of Eq.~\eqref{eq:phi_res_alpha_2} and Eq.~\eqref{eq:psd_res_2} can be obtained if not only one target, but a set of targets equaling the number of guide stars is considered ($\bmath{\alpha} = [\alpha_1, \alpha_2, ..., \alpha_{N_{GS}}]$). In this case, we can modify Eq.~\eqref{eq:phi_res_alpha_2} as:
\begin{equation}
    \label{eq:phi_res_alpha_3}
    \begin{split}
    \phi_{res}^{\bmath{\alpha}}(\nu) &= \phi_{turb}^{\bmath{\alpha}}(\nu) - H_{n,tomo}^{\bmath{\alpha}}(\nu) \big(\phi_{turb}^{\bmath{\theta_{GS}}}(\nu) + \phi_{n}(\nu) \big) \\
    &= Id \: \phi_{turb}^{\bmath{\alpha}}(\nu) - H_{n,tomo}^{\bmath{\alpha}}(\nu) \big(\phi_{turb}^{\bmath{\theta_{GS}}}(\nu) + \phi_{n}(\nu) \big)\\
    &= H_{r,tomo}^{\bmath{\alpha}}(\nu) \phi_{turb}^{\bmath{\alpha}}(\nu) \\
    & \:\:\: - H_{n,tomo}^{\bmath{\alpha}}(\nu) \big(\phi_{turb}^{\bmath{\theta_{GS}}}(\nu) - \phi_{turb}^{\bmath{\alpha}}(\nu) + \phi_{n}(\nu) \big) \, ,
    \end{split}
\end{equation}
where $H_{r,tomo}^{\bmath{\alpha}}$ is the tomographic RTF projected along $\bmath{\alpha}$, defined so that the relation $H_{r,tomo}^{\bmath{\alpha}}(\nu) + H_{n,tomo}^{\bmath{\alpha}}(\nu) = Id$ holds. This expression allows to differentiate the various contributions due to the rejection of turbulence (first term), to generalized anisoplanatism that is filtered as a noise by the AO loop (second plus third term) and to noise (last term). This is also shown by deriving the related temporal power spectrum:
\begin{equation}
    \label{eq:psd_res_3}
    \begin{split}
    S^{\bmath{\alpha}}_{res} (\nu) &= \: H_{r,tomo}^{\bmath{\alpha}}(\nu) S_{turb}^{\bmath{\alpha}}(\nu) H_{r,tomo}^{\bmath{\alpha} \: \dagger}(\nu) \\
    & \:\:\: + H_{n,tomo}^{\bmath{\alpha}}(\nu) S_n(\nu) H_{n,tomo}^{\bmath{\alpha} \: \dagger}(\nu) \\
    & \:\:\: + H_{n,tomo}^{\bmath{\alpha}}(\nu) \big(S_{turb}^{\bmath{\theta_{GS}}}(\nu) - S_{turb}^{\bmath{\alpha}}(\nu)\big) H_{n,tomo}^{\bmath{\alpha} \: \dagger}(\nu) \\
    & \:\:\: + 2 Re \Big[ H_{n,tomo}^{\bmath{\alpha}}(\nu) \big( S_{turb}^{\bmath{\alpha}} - S_{turb}^{\bmath{\theta_{GS}},\bmath{\alpha}} \big) \Big]\, ,
    \end{split}
\end{equation}
where the first and second term leads, respectively, to the temporal and noise error, while the remaining terms quantify the tomographic error as well as its temporal filtering by the MCAO loop.

\subsection{Pseudo-Open Loop control + MMSE reconstruction}
In the previous calculations, we considered a closed-loop control, that is, the reconstruction is performed on the residual measurements as shown in Eq.~\eqref{eq:phi_dm_1}. The reconstruction matrix $W$ is then intended as the pseudo-inverse of the projection matrix $P_{DM}^{\bmath{\theta_{GS}}}$, as derived in the Least Square Estimator (LSE) approach \citep{Madec99}. However, it has been demonstrated not to be the optimal approach to deal with the problem of badly and unseen modes \citep{Fusco01_1,Fusco01_2,Neichel09,Roux04} characterizing multi-conjugate adaptive optics correction and that the Minimum Mean Square Error (MMSE) approach can lead to better performance, even if compared to the Truncated LSE (TLSE) \citep{Pacheco04}. As the MMSE reconstructor operates on the pseudo-open loop measurements of the turbulent phase, it has to be included in a Pseudo-Open Loop control (POLC) \citep{Ellerbroek03}. In this context, we provide the expressions to derive the performance of MCAO systems also in the case of POLC and MMSE.

We modify Eq.~\eqref{eq:phi_dm_1} in order to consider a reconstruction acting on the pseudo-open loop measurements \citep{Basden19}:
\begin{equation}
    \label{eq:phi_dm_polc_1}
    \phi_{DM}(\nu) = H_{ol}(\nu) \big(W_{MMSE} \: \phi_{OL}^{\bmath{\theta_{GS}}}(\nu) - \phi_{DM}(\nu) \big) \, ,
\end{equation}
where $W_{MMSE}$ is the MMSE reconstructor and $\phi_{OL}^{\bmath{\theta_{GS}}}$ are the open-loop measurements that we write as:
\begin{equation}
    \label{eq:phi_open_loop}
    \phi_{OL}^{\bmath{\theta_{GS}}}(\nu) = \phi_{res}^{\bmath{\theta_{GS}}}(\nu) + \phi_{n}(\nu) + P_{DM}^{\bmath{\theta_{GS}}} \phi_{DM}(\nu) \, .
\end{equation}
We replace this expression in Eq.~\eqref{eq:phi_dm_polc_1}:
\begin{equation}
\begin{split}
    \label{eq:phi_dm_polc_2}
    \phi_{DM}(\nu) &= H_{ol}(\nu) \Big[W_{MMSE} \big( \phi_{res}^{\bmath{\theta_{GS}}}(\nu) + \phi_{n}(\nu) \\
    &\:\:\: + P_{DM}^{\bmath{\theta_{GS}}} \phi_{DM}(\nu) \big) - \phi_{DM}(\nu) \Big] \\
    &= H_{ol}(\nu) W_{MMSE} \big( \phi_{res}^{\bmath{\theta_{GS}}}(\nu) + \phi_{n}(\nu) \big) \\
    & \:\:\: + H_{ol}(\nu) (W_{MMSE} P_{DM}^{\bmath{\theta_{GS}}} - Id) \phi_{DM}(\nu) \, .
\end{split}
\end{equation}
We group the terms related to $\phi_{DM}$:
\begin{equation}
    \begin{split}
    \label{eq:phi_dm_polc_3}
        &\Big[ Id - H_{ol}(\nu) \big( W_{MMSE} \: P_{DM}^{\bmath{\theta_{GS}}} - Id \big) \Big] \phi_{DM}(\nu) \\
        &= H_{ol}(\nu) W_{MMSE} \big( \phi_{res}^{\bmath{\theta_{GS}}}(\nu) + \phi_{n}(\nu) \big)
        \, ,
    \end{split}
\end{equation}
and we obtain a final expression of the DMs phase:
\begin{equation}
\begin{split}
    \label{eq:phi_dm_polc_4}
    \phi_{DM}(\nu) &= \Big[ Id - H_{ol}(\nu) \big( W_{MMSE} \: P_{DM}^{\bmath{\theta_{GS}}} - Id \big) \Big]^{-1} \\
    &\:\:\: \times H_{ol}(\nu) \: W_{MMSE} \big( \phi_{res}^{\bmath{\theta_{GS}}}(\nu) + \phi_{n}(\nu) \big) \\
    &= \big[ Id + H_{ol}(\nu) K \big]^{-1} H_{ol}(\nu) \: W_{MMSE} \big( \phi_{res}^{\bmath{\theta_{GS}}}(\nu) + \phi_{n}(\nu) \big) \\
    &= H_{polc}(\nu) \: W_{MMSE} \big( \phi_{res}^{\bmath{\theta_{GS}}}(\nu) + \phi_{n}(\nu) \big) \, ,
\end{split}
\end{equation}
where we defined the matrices $K = Id - W_{MMSE} \: P_{DM}^{\bmath{\theta_{GS}}}$ and $H_{polc}=\big[ Id + H_{ol}(\nu) K \big]^{-1} H_{ol}(\nu)$.\newline
It follows that the results in Eqs.~\eqref{eq:phi_res_alpha_2} and \eqref{eq:psd_res_2} can still be used to compute the residual phase and PSD on target, but considering $H_{ol} = H_{polc}$ and $W = W_{MMSE}$ when taking into account POLC+MMSE.

\section{Tip-tilt anisoplanatism in MCAO-assisted astrometric observations}
\label{sec:aniso}
In this section, we use the formulation introduced in Sec.~\ref{sec:performance} as a tool to investigate the behavior of atmospheric tip-tilt residuals in MCAO-assisted observations and their impact on astrometric precision. Since, in the presented approach, the phase is intended as decomposed onto wavefront modes, we can derive the temporal PSD and the variance of tip-tilt residuals from Eq.~\eqref{eq:psd_res_2} and Eq.~\eqref{eq:var_res_1} respectively, by applying both equations to tip and tilt modes. 

\noindent Throughout the following analysis, we consider the contribution of all the modes to the turbulence-induced wavefront distortions and a reconstruction of tip-tilt at the ground and focus-astigmatisms at the high layer, based on the tip-tilt measurements from three NGSs in equilateral asterism. Such NGS loop can be used for the control of the \textit{null modes} \citep{Flicker03} in MCAO systems using a split tomography approach \citep{Gilles08}. The compensation for focus-astigmatisms at the pupil plane is not included in our configuration and this would provide an out of focus and astigmatic PSF; however, this is not a limitation for our analysis as we are interested in investigating the variations of tip-tilt in the field of view. As we do not consider the LGS-based correction of the higher orders, the results have to be intended as an upper limit to the atmospheric tip-tilt residuals. An extended study including the LGS loop will be the object of future works. We use an LSE reconstructor, as the control of modes up to the astigmatisms with a symmetric asterism and without noise does not foresee divergences in the system's behavior; thus, it does not require a threshold nor an MMSE reconstructor, as it would be expected in the real cases.

First, we analyse the dependence of on-axis tip-tilt residuals on the NGS asterism. Then, we introduce the contribution of the scientific integration time and, finally, we estimate relative tip-tilt residuals, that is the amount of differential tilt jitter error.

\subsection{On-axis tip-tilt residuals}
We consider the DM0 at 0m and the DM1 at 17km. We assume an equilateral asterism of NGSs centred at the origin of the field of view. We consider a 40-m telescope and the ELT median turbulence profile reported in \citet{elt_profile}, with a seeing of 0.644" and an average wind speed of 9.2 m/s. As we are mainly interested in the analysis of spatial anisoplanatism, we neglect the noise assuming NGSs with infinite flux. We also minimize the temporal error considering a loop with a frequency frame rate of 1kHz and where the control is a pure integrator with a delay given by the WFSs exposure time only. 

\begin{figure}
  \centering
  \includegraphics[width=\linewidth]{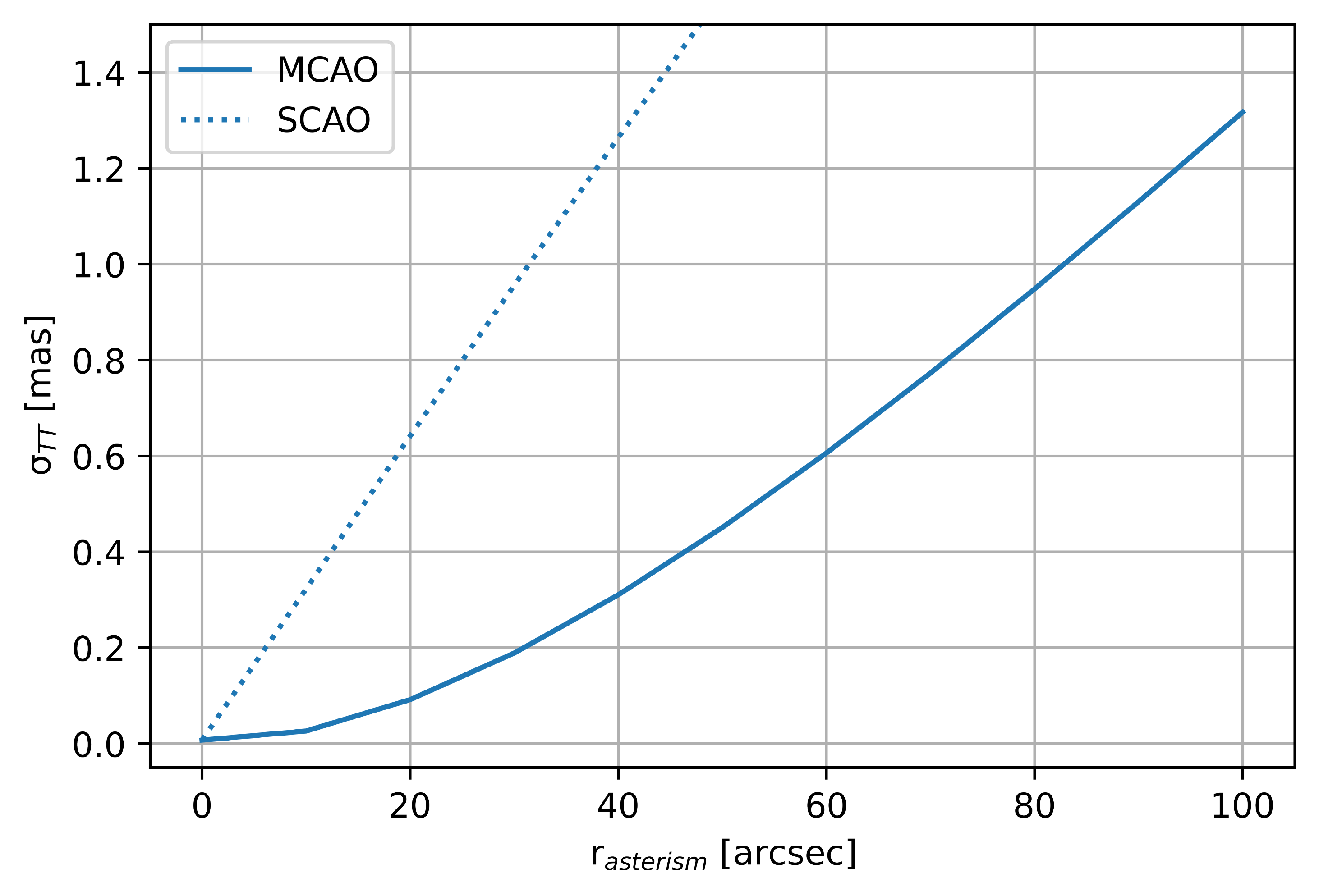}
  \caption{Tip-tilt residuals for a target at the origin of the field of view, as functions of the radius of the NGS asterism. The SCAO limit is also shown for comparison (dotted line); in this case, the values on the x-axis represent the angular separation between the target and the NGS.}
  \label{fig:tt_res_onaxis}
\end{figure}

In Fig.~\ref{fig:tt_res_onaxis}, we show the dependence on the asterism radius of tip-tilt residuals for a target on axis. The errors are computed from the integration of Eq.~\eqref{eq:psd_res_2}, applied to tip-tilt, over the temporal frequencies. The MCAO residuals are shown in comparison to the SCAO case, where the asterism radius becomes the angular separation of the NGS from the target; as expected from the larger isoplanatic patch provided by the MCAO correction, MCAO errors are reduced with respect to the SCAO ones. Moreover, we note that, differently from the SCAO case, whose errors linearly depend on the off-axis separation, MCAO residuals show a quadratic dependence on the NGSs separation. We can explain the different behaviors as follows: the turbulence-induced distortions that are observed on the pupil plane can be described by a combination of polynomials with increasing degree:
\begin{equation}
    \begin{split}
        \Delta x &= a_1 + a_2 x + a_3 y + a_4 x^2 + a_5 xy + a_6 y^2 + ...  \\
        \Delta y &= b_1 + b_2 y + b_3 x + b_4 y^2 + b_5 yx + b_6 x^2 + ... \, ,
    \end{split}   
\end{equation}
where the zeroth order coefficients ($a_1$, $b_1$) represent a global tip-tilt, that is a shift in $x$ and $y$ common to all directions of the field of view, the first order coefficients ($a_2$, $a_3$, $b_2$, $b_3$) represent the plate-scale distortions produced by the projection of focus and astigmatisms in altitude onto the tip-tilt in pupil, and so on for the higher orders. The covariance matrix of the distortions is $\langle \Delta \bmath{r} \Delta \bmath{r}^T \rangle$ , with $\Delta \bmath{r} = (\Delta x , \ \Delta y)$.
The SCAO, correcting with only a DM at the ground and using a single WFS, is able to compensate for the zeroth order of the distortions (i.e. overall pointing), leaving residual distortions that are then dominated by the first order (i.e. plate-scale variations). The MCAO, in our NGS-based configuration, removes a global tip-tilt with the DM0 and, in addition, is able to control the first order distortions by compensating for focus and astigmatisms with the DM1 conjugated in altitude. The residual distortions are, in this case, dominated by the second order. The sum of the diagonal terms of the residual distortions covariance matrix leads, for the SCAO case, to the following expression:
\begin{equation}
\begin{split}
    \sum_{i=1,2} \langle \Delta \bmath{r} \Delta \bmath{r}^T \rangle_{ii} &= (a_2 x + a_3 y + ...)^2 + (b_2 y + b_3 x + ...)^2 \\
    &= u (x^2 + y^2) + ... \, ,
\end{split}    
\end{equation}
and, for the MCAO case, to:
\begin{equation}
\begin{split}
    \sum_{i=1,2} \langle \Delta \bmath{r} \Delta \bmath{r}^T \rangle_{ii} &= (a_4 x^2 + a_5 xy + a_6 y^2 + ...)^2 \\
    & \:\:\: + (b_4 y^2 + b_5 yx + b_6 x^2 + ...)^2 \\
    &= v (x^2 + y^2)^2 + ... \, ,
\end{split}
\end{equation}
where the simplification in the coefficient $u$ for the former and $v$ for the latter is obtained by replacing the coefficients of the polynomial series with the proper coefficients that relate tip-tilt on the pupil plane with the higher orders on a meta-pupil in altitude (see Appendix \ref{sec:appendix}). If we consider ($x$, $y$) as the position of the target with respect to the NGS, we find a dependence of the variance on the second power of the separation for the SCAO case and on the fourth power for the MCAO case.

\begin{figure}
  \centering
  \includegraphics[width=\linewidth]{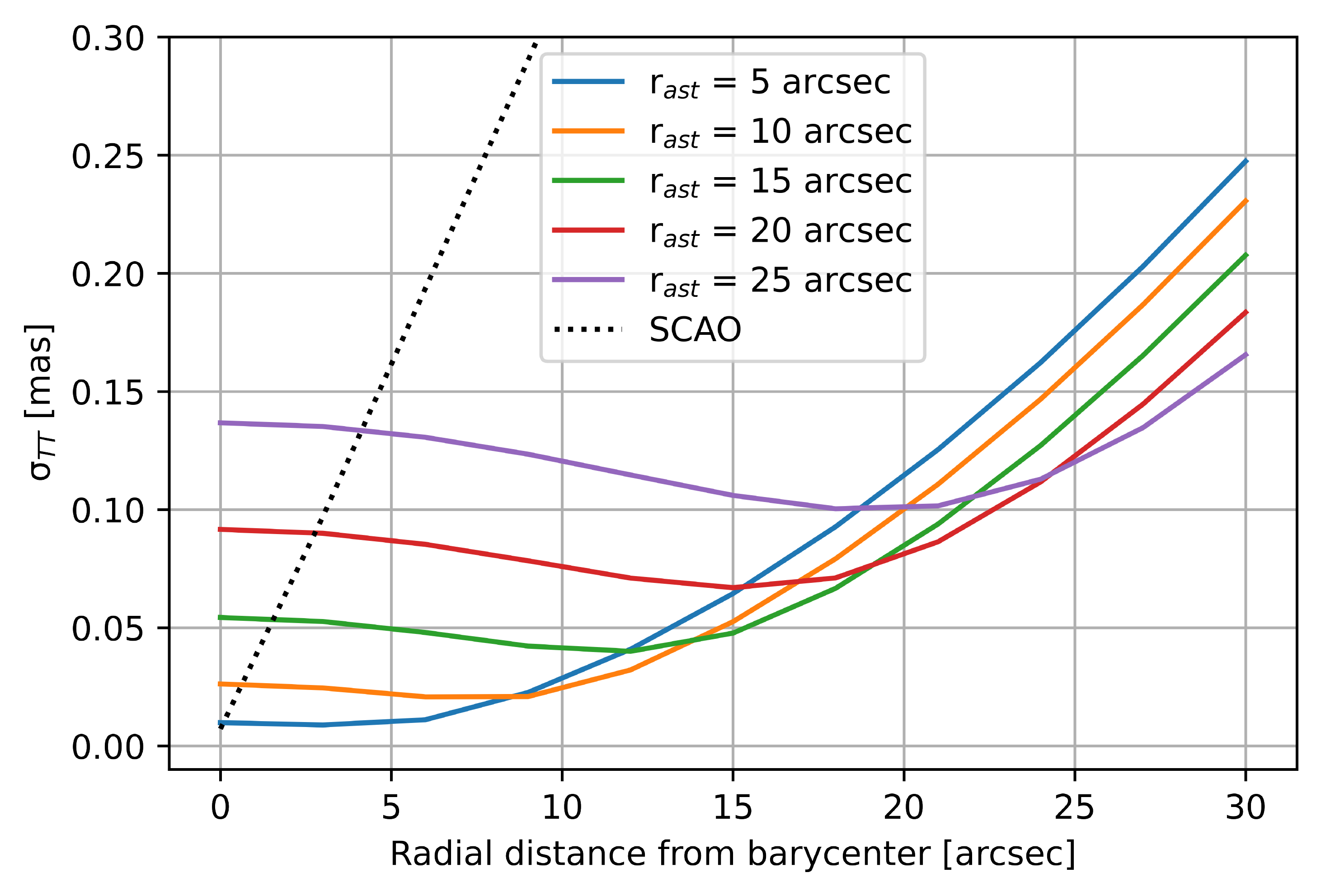}
  \caption{Tip-tilt residuals as functions of the target's radial distance with respect to the origin. The curves are shown for different values of the NGS asterism radius (r$_{ast}$) and the SCAO limit is also shown (dotted line).}
  \label{fig:tt_res_fov}
\end{figure}

In Fig.~\ref{fig:tt_res_fov}, we show the spatial distribution of tip-tilt residuals in the field of view. The errors are computed for targets at different radial separations from the origin (that also represents the barycenter of the asterism), the final values being obtained from the average over several polar angles in order not to be affected by the geometry of the asterism. 
The errors show similar values for targets within the NGS asterism and increase outside of the asterism, where tip-tilt is indeed not controlled. The minimum of the curves is not exactly at a distance equal to the asterism radius value, depending on the fact that the targets at an angular separation equal to the asterism radius fall outside of the NGSs triangle (except the ones with the same exact polar angles as the NGSs ones), where tip-tilt is worse controlled.

\subsection{Scaling of tip-tilt residuals with the scientific integration time}
\label{sec:sci_time}
The previous results, obtained from a pure integration of Eq.~\eqref{eq:psd_res_2}, represent the case where the fluctuations in position due to tip-tilt residuals are fully integrated within the exposure and thus impact entirely on the shape and size of the PSF, leading to the PSF elongation. This effect contributes to the astrometric error due to photon noise \citep{Lindegren78}:
\begin{equation}
   \sigma \sim \frac{FWHM}{SNR} \, , 
\end{equation}
where $FWHM$ is the full width at half maximum of the PSF and $SNR$ is the signal-to-noise ratio. Regardless of the residual value contributing to the FWHM, this source of error can be ideally reduced to zero if we assume a source with infinite SNR. In this case, tip-tilt residuals would not affect the astrometric precision. On the other hand, if tip-tilt residuals are not fully integrated within the exposure, fluctuations in position due to the residual jitter are observed between successive frames, these affecting astrometric precision despite the source flux. Thanks to the knowledge of the temporal PSD of the residuals, we can analytically describe the residual jitter between successive frames by still following an approach that makes use of temporal transfer functions, as in Sec.~\ref{sec:performance}.

We write the expression of the phase residuals that are left after a scientific integration of length $T$ as:
\begin{equation}
    \label{eq:phi_sci_freq_1}
    \phi_{res, T}^{\alpha}(\nu) = H_{T}(\nu) \phi_{res}^{\alpha}(\nu) \, ,
\end{equation}
where $\phi_{res}^{\alpha}$ is given by Eq.~\eqref{eq:phi_res_alpha_2} and $H_{T}$ is the temporal transfer function of the scientific camera, that is the Laplace or $Z$-transform of the time-average operation. In the Laplace case, the expression is given by:
\begin{equation}
    \label{eq:camera_tf}
    \begin{split}
    H_{T}(\nu) &= \dfrac{1}{T} \widetilde{\Pi}_{T}(\nu) \\
    &= sinc(\pi \nu T) e^{-i \pi \nu T} \, ,
    \end{split}
\end{equation}
where $\widetilde{\Pi}_{T}$ denotes the transform of the rectangular function $\Pi_{T}$. From Eq.~\eqref{eq:psd_res_1} and Eq.~\eqref{eq:phi_sci_freq_1}, we can get the expression of the residual PSD for scientific frames of length $T$:
\begin{equation}
    \label{eq:psd_sci_1}
    S_{res, T}^{\alpha}(\nu) = |H_{T}(\nu)|^2 S_{res}^{\alpha}(\nu) \, ,
\end{equation}
where $S_{res}^{\alpha}$ is given by Eq.~\eqref{eq:psd_res_2}.

\begin{figure}
    \centering
    \includegraphics[width=\linewidth]{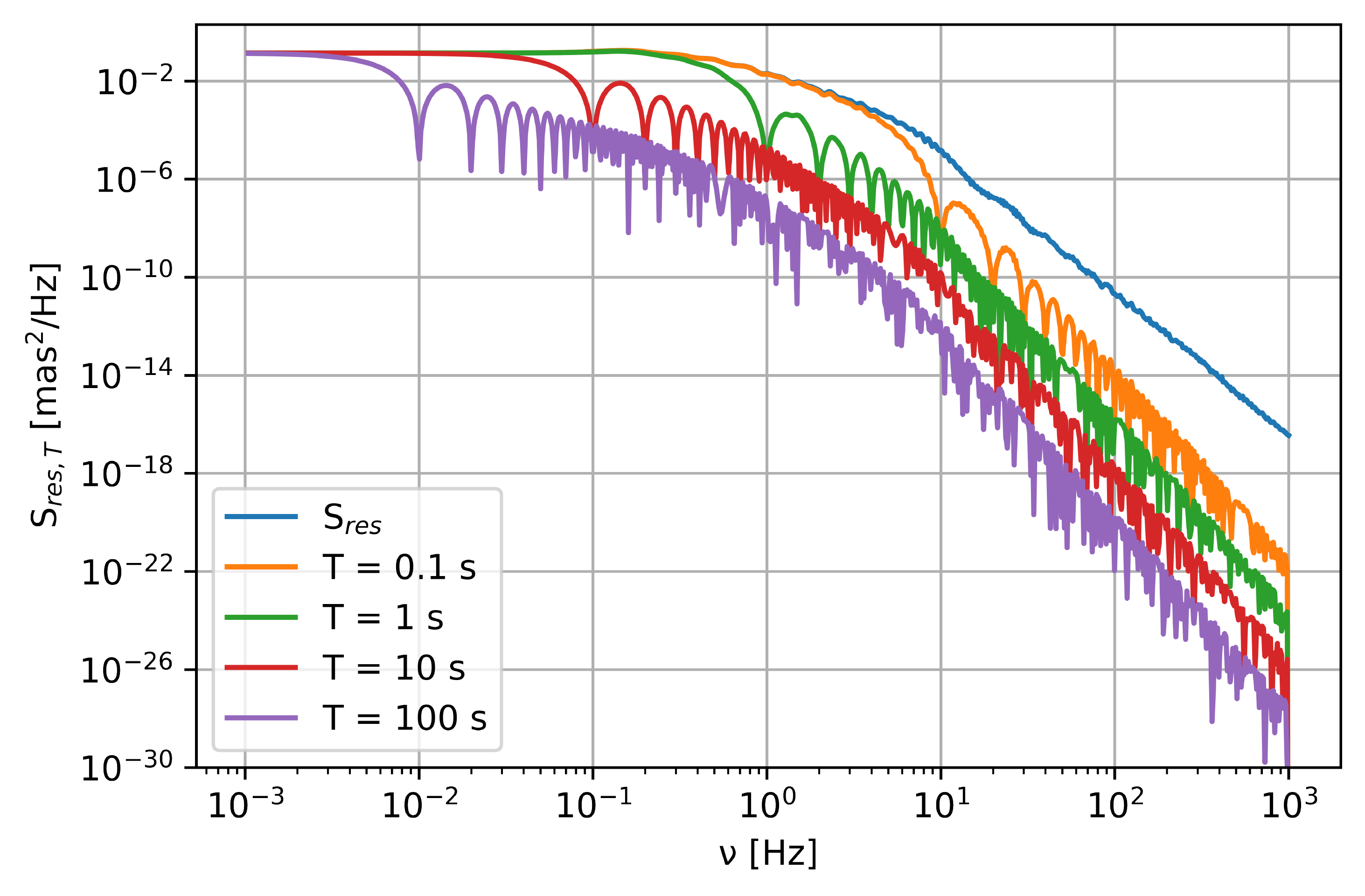}
    \caption{Temporal power spectrum of the residual jitter between successive frames, for scientific exposures of 0.1s (orange), 1s (green), 10s (red), 100s (purple). The configuration is the same as Fig.~\ref{fig:tt_res_onaxis}, with a target on axis and an asterism radius of 40". The unaveraged tip-tilt residual PSD is also shown for comparison (blue).}
    \label{fig:sci_psds}
\end{figure}

\begin{figure}
    \centering
    \includegraphics[width=\linewidth]{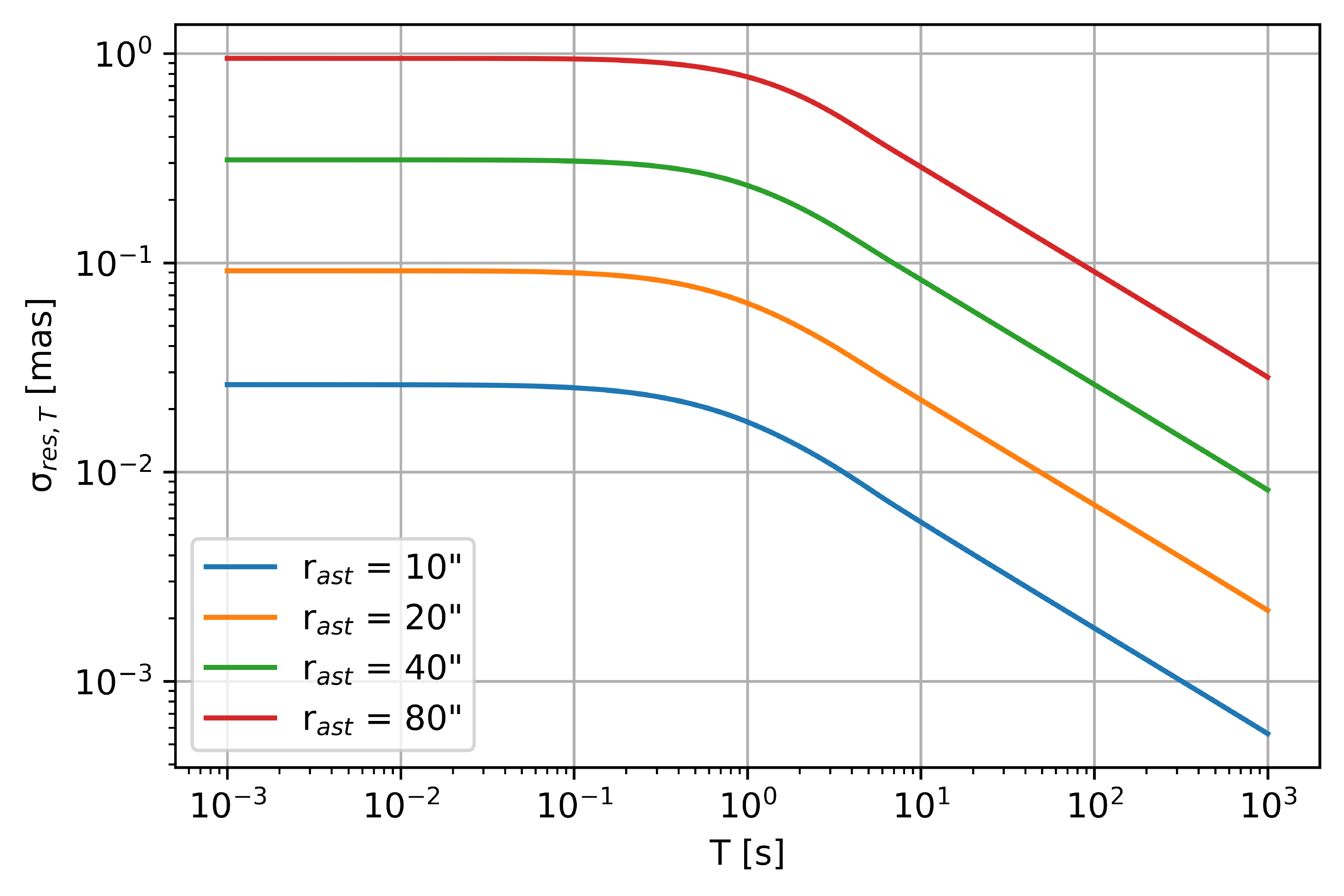}
    \caption{Tip-tilt residual error on axis as a function of the scientific integration time. The configuration is the same as Fig.~\ref{fig:sci_psds}, with the asterism radius varying from 10" to 80".}
    \label{fig:sci_std_vs_T}
\end{figure}

The results of this expression, as applied to tip and tilt, are shown in Fig.~\ref{fig:sci_psds}, where on-axis tip-tilt residual PSDs are plotted for different integration times.
The impact of the scientific exposure depends on the relation between the cut-off frequency of the camera transfer function and the one of the residual PSD. The camera transfer function acts as a low-pass filter with a cut-off frequency $\nu_{H_{sci}} = 1/T$. If $\nu_{H_{sci}}$ is either larger or about the same as the tip-tilt residual PSD cut-off frequency ($\nu_{S_{res}} \simeq 0.6 \: v/D$, with $v$ the wind velocity and $D$ the telescope diameter \citealt{Conan95}), the scientific integration is not long enough to average the residuals and the position jitter observed between different exposures is emphasized. Indeed, in this case, the camera is either unable to filter any frequency of the PSD, or it filters only the frequencies that are larger than $\nu_{S_{res}}$, where the energy falls rapidly to zero.
As the integration time increases, $\nu_{H_{sci}}$ becomes smaller than $\nu_{S_{res}}$ and the camera transfer function passes the frequencies where the PSD is flat, leaving then a residual variance that is proportional to $1/T$. Thus, the root-mean-square (RMS) is proportional to $T^{-1/2}$. This behavior is shown in Fig.~\ref{fig:sci_std_vs_T}: for integration times smaller than the inverse of $\nu_{S_{res}}$, tip-tilt residuals do not depend on $T$ and the curve is flat, while it follows a $T^{-1/2}$ law for larger times. The $T^{-1/2}$ power law is in agreement with the assumptions and the results that are present in the literature \citep{Ammons11,Cameron09,Ellerbroek07}.

\subsection{Differential tilt jitter}
\label{sec:dtj}
The results in Sec.~\ref{sec:sci_time} give information about the repeatability of the position measurement of a single source. However, the science cases of future instruments show a major interest in relative astrometry, that is, in measuring the distance between sources. In order to be able to estimate the precision in the distance measurements, we extend the analysis to differential tilt jitter. This effect is well known for SCAO systems but, to our knowledge, is less well understood and no expression is present in the literature to compute this error for MCAO systems. In this context, we present an analytical expression for this flavour of adaptive optics as well, by using the results in Sec.~\ref{sec:performance}.

We consider two sources in directions $\alpha$ and $\beta$ and we describe the differential jitter phase through the difference between the residual phases in the two directions:
\begin{equation}
    \label{eq:phi_dtj_1}
    \phi^{\alpha,\beta}_{DTJ}(\nu) = \phi_{res}^{\alpha}(\nu) - \phi_{res}^{\beta}(\nu) \, .
\end{equation}
The temporal PSD is then:
\begin{equation}
    \label{eq:psd_dtj_1}
    \begin{split}
    S_{DTJ}^{\alpha,\beta}(\nu) &= \Big\langle \phi^{\alpha,\beta}_{DTJ}(\nu) \: \phi^{\alpha,\beta \: \dagger}_{DTJ}(\nu) \Big\rangle\\
    &= \Big\langle \big(\phi_{res}^{\alpha}(\nu) - \phi_{res}^{\beta}(\nu)\big) \big(\phi_{res}^{\alpha}(\nu) - \phi_{res}^{\beta}(\nu)\big)^{\dagger} \Big\rangle \, .
    \end{split}
\end{equation}

For SCAO systems, the difference between residual phases simplifies into the difference between turbulent phases because, as already pointed out in Sec.~\ref{sec:performance}, the correction phase is common to all directions. The reasoning leads to the following expression of differential tilt jitter PSD for the SCAO case:
\begin{equation}
    \label{eq:psd_dtj_scao}
    S_{DTJ}^{\alpha,\beta}(\nu) = 2\big(S_{turb}(\nu) - S_{turb}^{\alpha,\beta}(\nu) \big) \, ,
\end{equation}
where $S_{turb}^{\alpha,\beta}$ is the CPSD of turbulence between the two directions and where we considered $S_{turb} = S_{turb}^{\alpha} = S_{turb}^{\beta}$, having assumed a homogeneous and isotropic turbulence. The expression (integrated over the temporal frequencies) is in agreement with the results that are present in the literature \citep{Sandler94,Clenet15}.

\begin{figure}
  \centering
  \includegraphics[width=\linewidth]{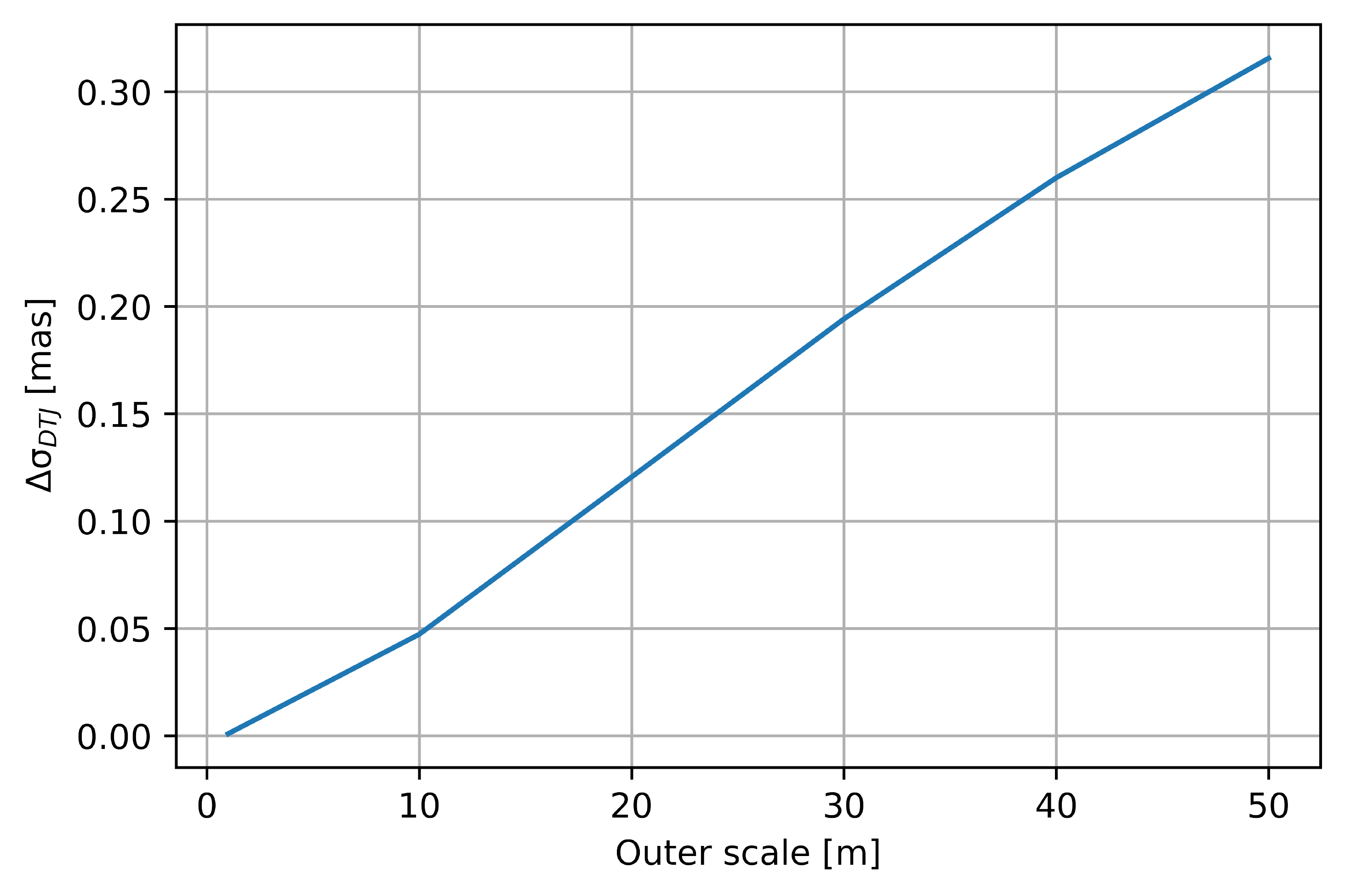}
  \caption{Difference between SCAO and MCAO differential tilt jitter error ($\Delta\sigma_{DTJ}$ = $\sigma_{DTJ,SCAO}$ - $\sigma_{DTJ,MCAO}$) as a function of the outer scale. The telescope, DMs, NGSs, turbulence configurations are the same as in Fig.~\ref{fig:tt_res_onaxis}. The targets' angular separation is 5" and the asterism radius is 40".}
  \label{fig:dtj_vs_outerscale}
 \end{figure}

\begin{figure}
  \centering
  \includegraphics[width=\linewidth]{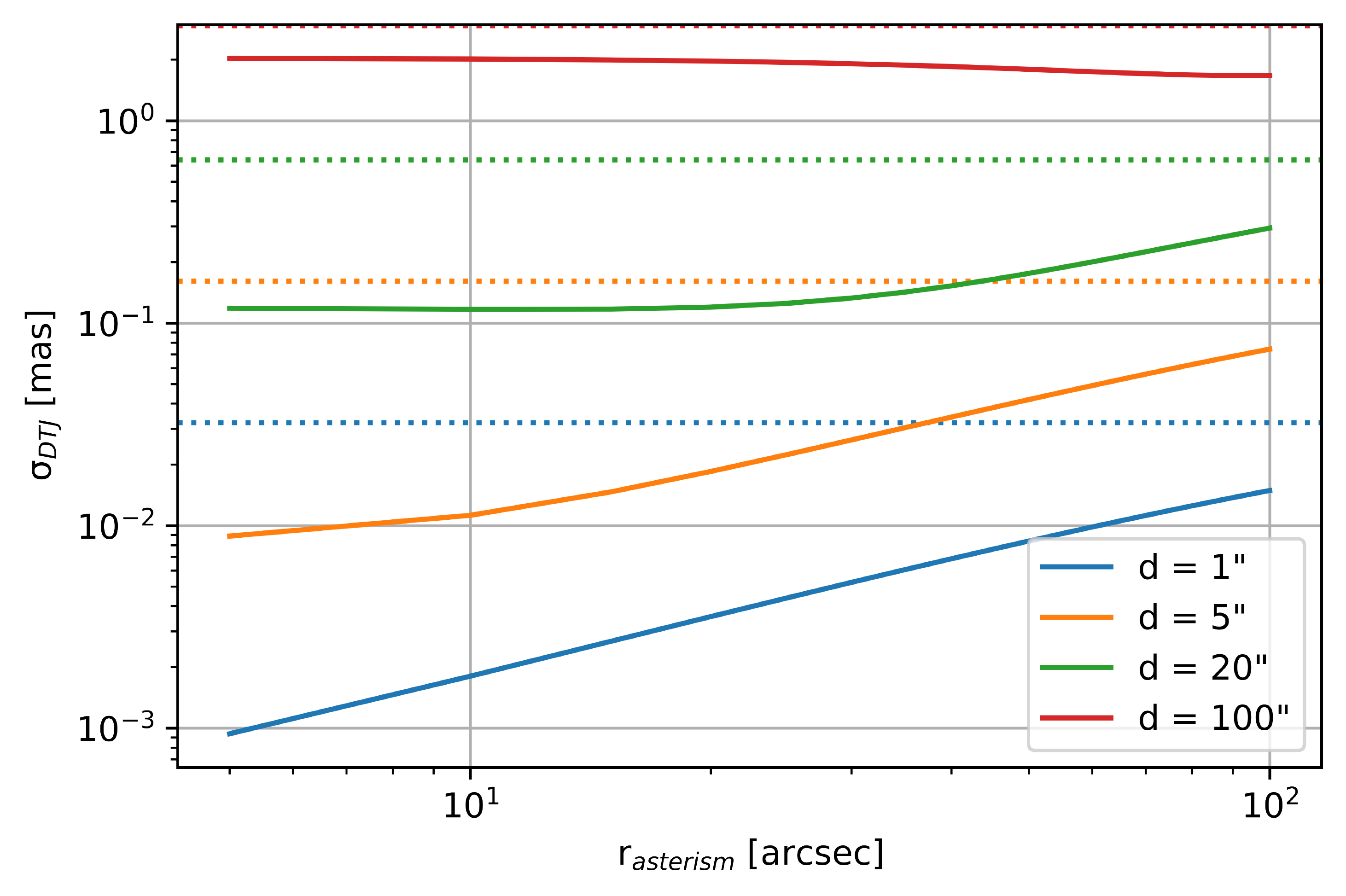}
  \caption{MCAO differential tilt jitter error as a function of the NGS asterism radius. The colors show different values of the distance between the astrometric targets. For each curve, the SCAO case is shown as comparison (dotted lines).}
  \label{fig:dtj_vs_asterism_radius}
\end{figure}

For MCAO systems, we can replace $\phi_{res}^{\alpha}$ and $\phi_{res}^{\beta}$ with the expression in Eq.~\eqref{eq:phi_res_alpha_2} applied to $\alpha$ and $\beta$ respectively. We obtain:
\begin{equation}
    \label{eq:psd_dtj_2}
    \begin{split}
    S_{DTJ}^{\alpha,\beta}(\nu) &= \: 2\big(S_{turb}(\nu) - S_{turb}^{\alpha,\beta}(\nu) \big) \\
    & \:\:\: + \Delta H_{n,tomo}^{\alpha,\beta}(\nu) \big(S_{turb}^{\bmath{\theta_{GS}}}(\nu) + S_{noise}(\nu) \big)\Delta H_{n,tomo}^{\alpha,\beta}(\nu)^{\dagger} \\
        & \:\:\: - 2 Re \Big[ \Delta H_{n,tomo}^{\alpha,\beta}(\nu) \big(S_{turb}^{{\bmath{\theta_{GS}}},\alpha}(\nu) - S_{turb}^{{\bmath{\theta_{GS}}},\beta}(\nu)\big) \Big] \, ,
    \end{split}
\end{equation}
where we defined $\Delta H_{n,tomo}^{\alpha,\beta}(\nu) = H_{n,tomo}^{\alpha}(\nu) - H_{n,tomo}^{\beta}(\nu)$. It is worth noting that, if taking the SCAO limit of this expression, we get $H_{n,tomo}^{\alpha} = H_{n,tomo}^{\beta}$ and we retrieve the results in Eq.~\eqref{eq:psd_dtj_scao}. Equation \eqref{eq:psd_dtj_2} shows that differential tilt jitter error in MCAO systems is given by the SCAO case error (first two terms) and additional terms depending on the correction (asterism/targets geometry, temporal filtering of the AO loop, noise) and on spatiotemporal cross-correlations of the turbulence. These additional terms might reduce the error with respect to the SCAO case, as shown in Fig.~\ref{fig:dtj_vs_outerscale} and Fig.~\ref{fig:dtj_vs_asterism_radius}. 
In the former, the RMS of the difference between the variances obtained from Eq.~\eqref{eq:psd_dtj_scao} and Eq.~\eqref{eq:psd_dtj_2} as a function of the outer scale is plotted. As expected, the discrepancy between the SCAO and MCAO values increases with the outer scale, as a larger outer scale leads to larger cross-correlations that help reduce the differential tilt jitter error in the MCAO correction. In the latter, the MCAO differential tilt jitter error as a function of the NGS asterism radius is shown. The smaller cross-correlations given by larger asterisms determine an increasing of the differential tilt jitter error with the asterism radius. This is evident when the distance is small and both targets are included within the asterism (d = 1", 5"); for larger distances, the errors are about constant up to an asterism radius comparable to the targets' separation and then show the increasing behavior.

As in Sec.~\ref{sec:sci_time}, we can also take into account the contribution of the scientific exposure on the differential tilt jitter error, through the temporal filtering of the camera integrating over $T$:
\begin{equation}
    \label{eq:phi_dtj_sci}
    \phi^{\alpha,\beta}_{DTJ, T}(\nu) = H_{T}(\nu) \big(\phi_{res}^{\alpha}(\nu) - \phi_{res}^{\beta}(\nu)\big) \, .
\end{equation}
The PSD of time-averaged differential tilt jitter is then:
\begin{equation}
    \label{eq:psd_dtj_sci}
    S^{\alpha,\beta}_{DTJ, T}(\nu) = |H_{T}(\nu)|^2 S^{\alpha,\beta}_{DTJ}(\nu) \, ,
\end{equation}
where $S^{\alpha,\beta}_{DTJ}$ is given by Eq.~\eqref{eq:psd_dtj_scao} for SCAO and by Eq.~\eqref{eq:psd_dtj_2} for MCAO.

\section{Application: Differential tilt jitter error for MAVIS and MAORY}
\label{sec:application}
In this section, we use Eq.~\eqref{eq:psd_dtj_2} to investigate the contribution of differential tilt jitter error on the future astrometric observations; as case studies, we consider MAVIS at the VLT and MAORY at the ELT. 

\noindent In Table~\ref{table:mav_mao_parameters}, we summarize the main parameters that we used to describe the two systems. The maximum value of the asterism radius represents the technical field of view (120" for MAVIS and 160" for MAORY). 
\begin{table}
\centering
\begin{tabular}{ |c|c|c| } 
 \hline
 \textit{} & MAVIS & MAORY \\
 \hline
 $D$ [m] & 8 & 39 \\
 $h_{DM_0}$ [m] & 0 & 600 \\ 
 $h_{DM_1}$ [m] & 13500 & 17000 \\
 $r_{asterism}$ ["] & 10, 30, 50, 60 & 30, 55, 70, 80 \\
 $r_{FoV}$ ["] & 15 & 30 \\
 \hline
\end{tabular}
\vspace{.2cm}
\caption{Telescope diameter, DMs conjugation height, set of asterism radii and scientific field of view radius used to derive the differential tilt jitter error for MAVIS- and MAORY-assisted observations. The outer scale used for both cases is 25m.}
\label{table:mav_mao_parameters}
\end{table}
\noindent As in Sec.~\ref{sec:aniso}, we assume equilateral asterisms of NGS with infinite flux in order to neglect the contribution of noise. The measurements from the three NGSs allow to reconstruct tip and tilt, that are corrected on the DM0, and focus-astigmatisms, applied on the DM1. We consider a closed loop, where the control is a pure integrator working at 1kHz and where we minimize the latency by considering a delay due to the WFSs integration time only. For the computation of the PSDs and CPSDs of turbulence, we used the same turbulence profile as in Sec.~\ref{sec:aniso}, with a zenith angle of 30$^{\circ}$. 

\begin{figure}
\centering
\begin{minipage}{.45\textwidth}
  \centering
  \includegraphics[width=\linewidth]{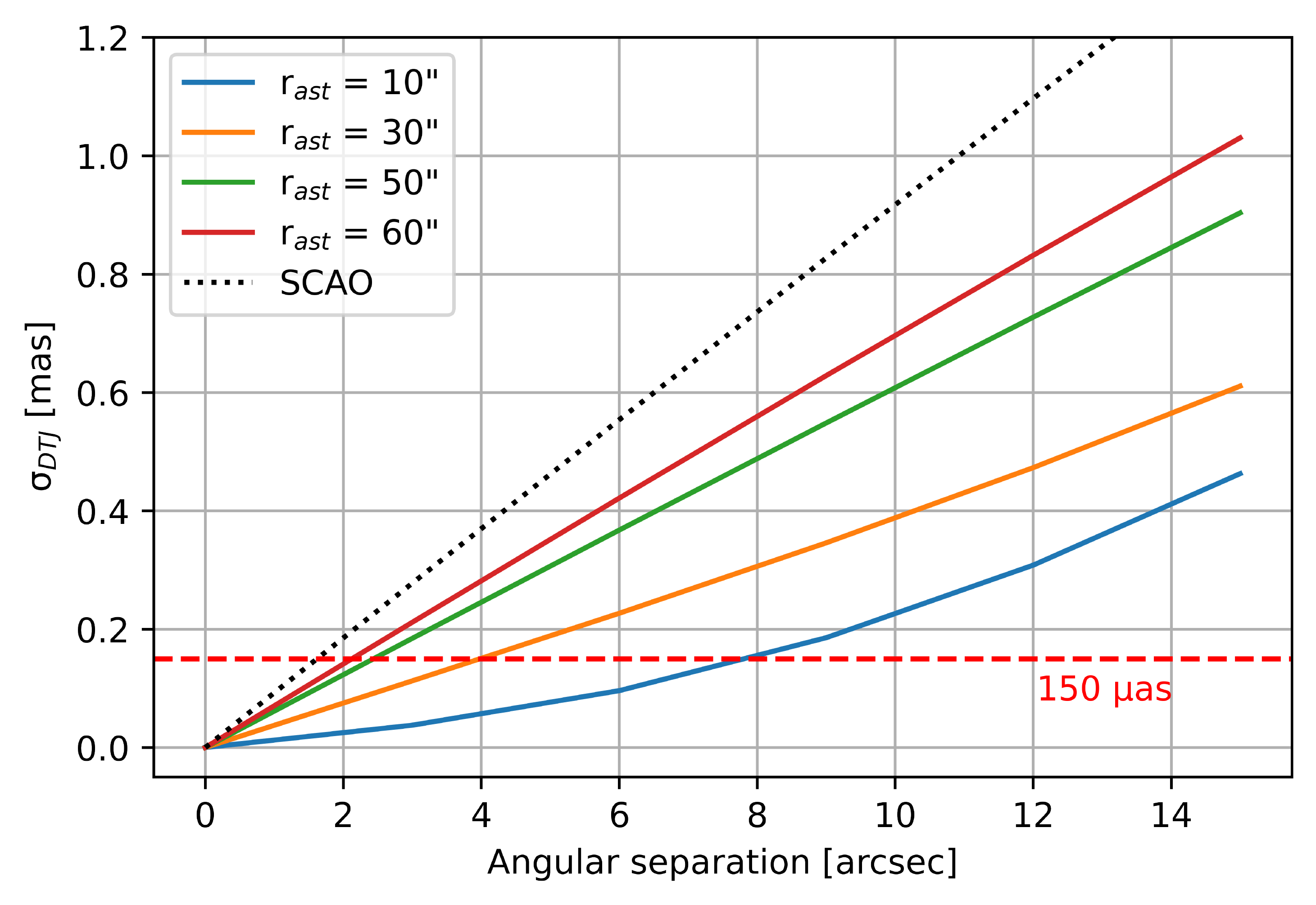}
\end{minipage}\hfill
\begin{minipage}{.45\textwidth}
  \centering
  \includegraphics[width=\linewidth]{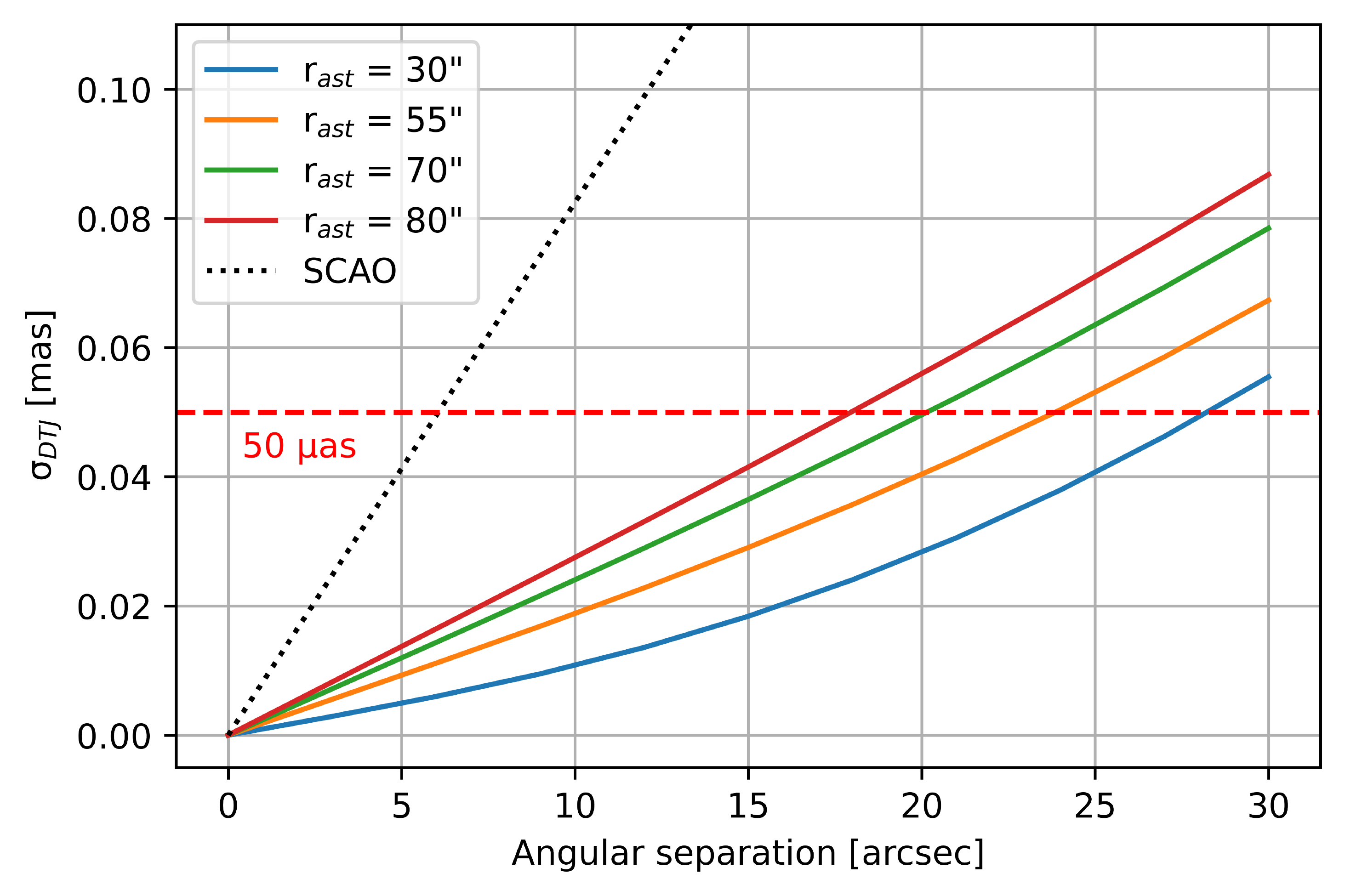}
\end{minipage}
\caption{Differential tilt jitter error as a function of the angular separation for MAVIS- (top) and MAORY- (bottom) assisted observations of length $T$=30s. The SCAO case (dotted black line) is shown for comparison, as well as the astrometric precision requirement (dashed red line) of the two systems.}
\label{fig:dtj_mao_mav}
\end{figure}

\noindent In Fig.~\ref{fig:dtj_mao_mav}, we show the differential tilt jitter error for MAVIS and MAORY, obtained for typical scientific exposures of $T$=30s. The error is computed considering the first source at the origin of the field of view and varying the distance of the second source up to the edge of the scientific field of view.

In order not to be affected by the geometry of the asterism of NGSs, for each separation we made an azimuthal average of the errors obtained at different polar coordinates. The plots show that differential tilt jitter can introduce errors on relative astrometry up to $\sim$0.4-1mas for MAVIS and $\sim$60-90µas for MAORY at the edge of the field of view. As shown in Sec.~\ref{sec:sci_time}, this source of error can be reduced with the integration time; if the measurements, for instance, can be averaged over $\sim$30 minutes of exposures, the relative astrometric error due to differential tilt jitter is reduced by a factor of $\sim$ 8 and becomes smaller than the requirement value over the whole field of view for both cases. 

\begin{figure}
\centering
\begin{minipage}{.45\textwidth}
  \centering
  \includegraphics[width=\linewidth]{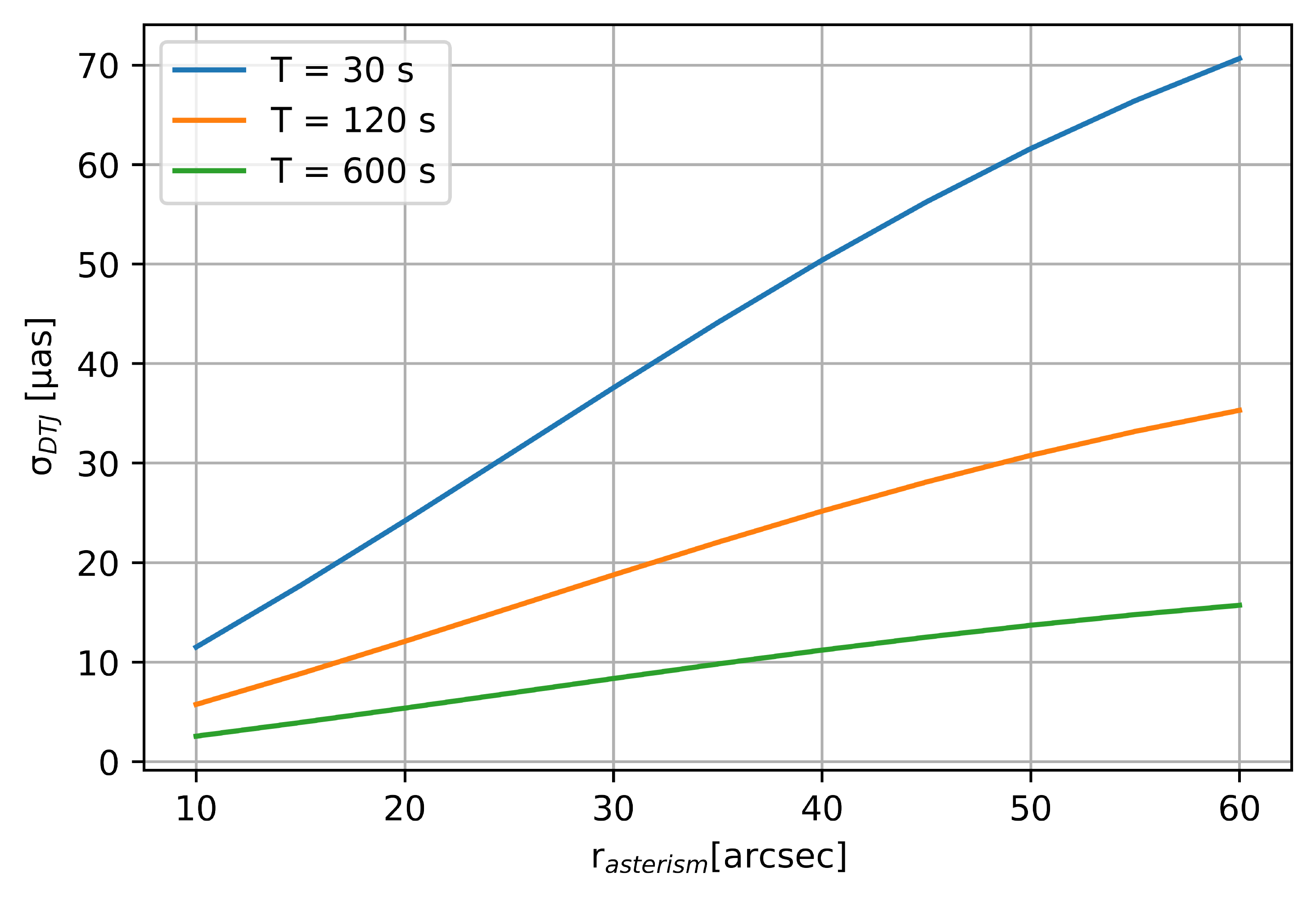}
\end{minipage}\hfill
\begin{minipage}{.45\textwidth}
  \centering
  \includegraphics[width=\linewidth]{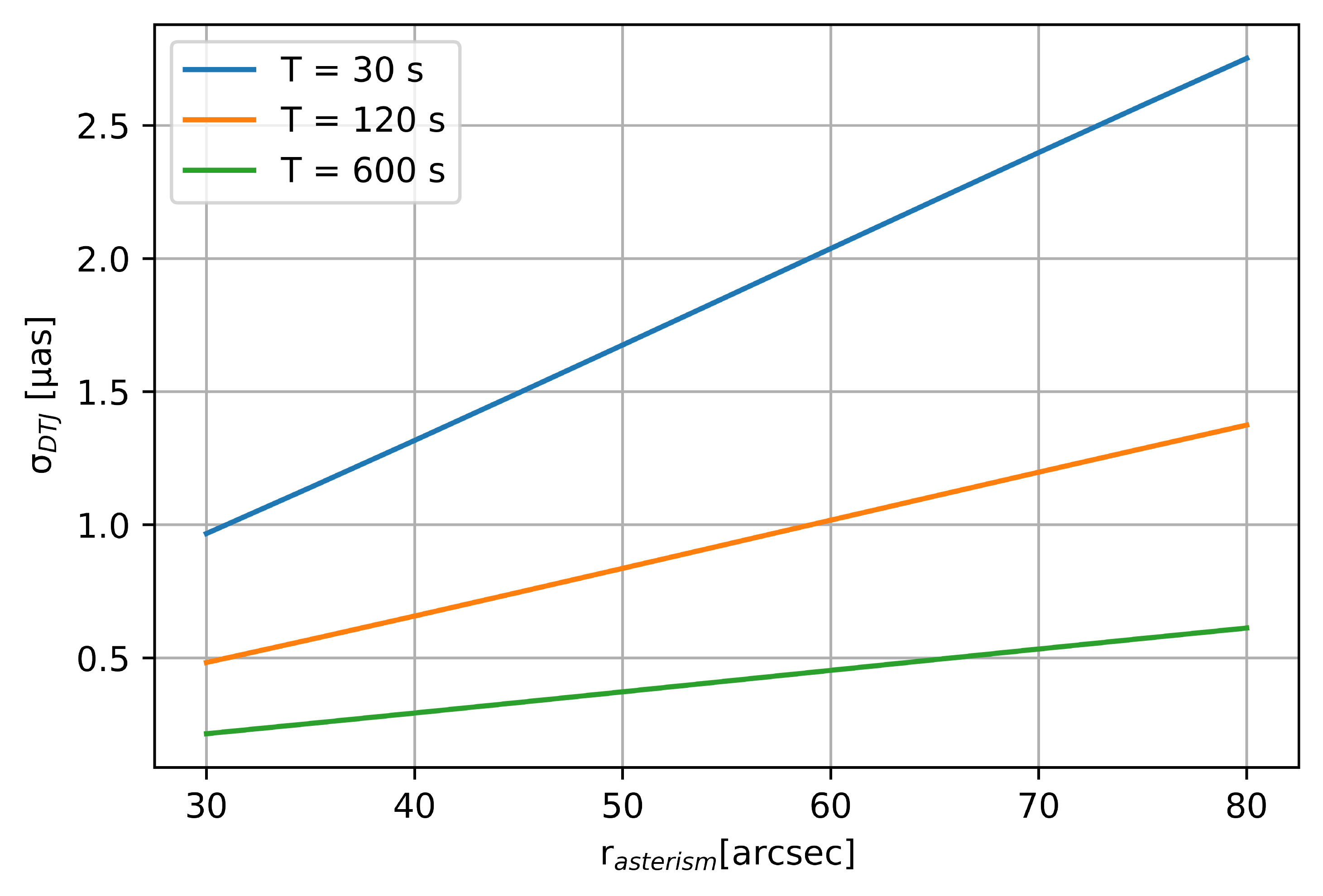}
\end{minipage}
\caption{Differential tilt jitter error for targets separation of 1" as a function of the NGS asterism radius for MAVIS- (top) and MAORY- (bottom) assisted observations. The results are plotted for $T=30, 120$ and $600$s and show the scaling with $T^{-1/2}$ that has been demonstrated in Sec.~\ref{sec:sci_time}.}
\label{fig:dtj_mao_mav_1arcsec}
\end{figure}

\noindent Current specifications suggest a major interest in high precision relative astrometry for separations up to 1" \citep{Rigaut20}. For a better visualization of this scale, in Fig.~\ref{fig:dtj_mao_mav_1arcsec} we show the differential tilt jitter error as a function of the asterism radius for a fixed distance of 1". The plots show that differential tilt jitter error should not represent a relevant contribution to the MAORY astrometric error budget for these separations, even considering the goal of 10µas. For MAVIS, the error shows to be within the requirement of 150µas, but not compliant with the goal of 50µas for asterisms with radius larger than 40" and for the typical exposure time of 30s. In this case, the possibility to average over longer integration times is required.

\noindent It is worth pointing out that these results show the contribution of atmospheric tip-tilt residuals in terms of differential tilt jitter only. The contribution of tip-tilt residuals on the astrometric error in terms of the centroiding error is not considered (that is equivalent to assume targets with infinite SNR). Moreover, the contribution of temporal errors of the AO loop is minimized and noise terms are neglected. On the other hand, it should be considered that the differential tilt jitter error could be calibrated out through dedicated coordinate transforms, if reference sources are available in the field \citep{Fritz10,Cameron09}. We also expect the error to be reduced if an LGS loop controlling the higher orders than the astigmatisms is included. In this context, these results have to be considered as an upper limit. An extended study about the impact of the LGS loop residuals on the tip-tilt modes is intended to be the object of future works.

\section{Conclusion}
We have presented an analytical formalism to derive the temporal PSD of the wavefront residuals of an MCAO correction. The formulation includes tomographic, noise and temporal errors. The general framework allows to select the telescope diameter, the asterism of either NGSs or LGSs, the DMs configuration, the turbulence profile and the modes of distortion that are sensed through the GSs and compensated by the DMs. We derived an expression for both a closed loop control with an LSE reconstruction and a pseudo-open loop control with an MMSE reconstruction. We applied the results to an NGS-based MCAO configuration in order to analyse the spatial and temporal behavior of tip-tilt residuals: we found a quadratic dependence of the on-axis residuals on the angular separation of the asterism, that we demonstrated to be consistent with the control of plate-scale distortions operated by the MCAO correction; we also verified the scaling of the residuals with the square root of the scientific exposure time by means of the temporal transfer function of the scientific camera. We analysed differential residuals as well and we provided an analytical expression for the differential tilt jitter error. We showed that the cross-correlations between the GSs of the asterism and between the GSs and the targets play a role in reducing this source of error with respect to the SCAO case and that parameters like the outer scale and the radius of the asterism can be crucial to properly decrease the differential tilt jitter in MCAO systems. Though these parameters are not under control, it is worth considering them during the preparation of astrometric observations. We finally used our results to quantify the contribution of the differential tilt jitter error to the future astrometric observations, choosing MAORY and MAVIS as case studies. In the case of equilateral asterism of NGSs and considering the possibility of averaging over several exposures, differential tilt jitter should not be the dominant limiting factor to the astrometric precision of these systems.

\section*{Acknowledgements}
The authors thank Carmelo Arcidiacono for fruitful discussion.
This work has been partially funded by ADONI - the ADaptive Optics National laboratory of Italy.

\section*{Data Availability}
No new data were generated or analysed in support of this research.


\bibliographystyle{mnras}
\bibliography{mcao_res}



\appendix
\section{Computation of residual distortions on a pupil plane}
\label{sec:appendix}
We use the same notation as \citet{Flicker02} to relate tip-tilt on the telescope pupil plane with the distortions on a layer of turbulence at altitude $h_l$. The phase observed at the pupil plane can be seen as a linear combination of tip and tilt:
\begin{equation}
\varphi(\bmath{x}, \bmath{\theta}, t) = \sum_{k=2}^3 \gamma_k(\bmath{\theta}, t) Z_k(\bmath{x}/R) \, ,
\end{equation}
where $\varphi$ is the phase observed at coordinates $\bmath{x}$ on the pupil plane for a source at position $\bmath{\theta}$, $R$ is the telescope pupil radius, $Z_k$ is the kth Zernike mode and $\gamma_k(\bmath{\theta}, t)$ are time and field dependent coefficients relating tip-tilt on the pupil plane with all the modes of distortion on a meta-pupil in altitude:
\begin{equation}
    \gamma_k(\bmath{\theta}, t) = \sum_{i=2}^N c_{ik}(\bmath{\theta})A_i(t) \, ,
\end{equation}
where the coefficients $c_{ik}(\bmath{\theta})$ are defined as (e.g. \citealt{Negro84}):
\begin{equation}
    \begin{split}
        \bmath{c_{2}}(\bmath{\theta}) &= [1, 0, 2\sqrt{3}\theta_x, \sqrt{6}\theta_y, \sqrt{6}\theta_x, 6\sqrt{2}\theta_x\theta_y, \\
        & \:\:\:\:\:\:\: 3\sqrt{2}(3\theta_x^2 + \theta_y^2), 6\sqrt{2}\theta_x\theta_y, 3\sqrt{2}(\theta_x^2 - \theta_y^2), \cdots] \\
        \bmath{c_{3}}(\bmath{\theta}) &= [0, 1, 2\sqrt{3}\theta_y, \sqrt{6}\theta_x, -\sqrt{6}\theta_y, 3\sqrt{2}(\theta_x^2 + 3\theta_y^2), \\
        & \:\:\:\:\:\:\: 6\sqrt{2}\theta_x\theta_y, 3\sqrt{2}(\theta_x^2 - \theta_y^2), - 6\sqrt{2}\theta_x\theta_y, \cdots] \, 
    \end{split}
\end{equation}
and $A_i$ as:
\begin{equation}
        A_{[2,3]}(t) = a_{[2,3]l}(t) R/R_l; \:\: A_{[4:10]}(t) = a_{[4:10]l}(t) h_l R/R_l^2; \:\: \cdots \, 
\end{equation}
Due to the orthogonality of the Zernike, the phase variance can be computed as:
\begin{equation}
    \begin{split}
        \sigma_{\varphi}^2 &= tr(C_{\gamma}) = \sum_{k=2}^3 C_{\gamma}^{kk} \\
        &= \sum_{k=2}^3 \bigg \langle \Big(\sum_{i=2}^N c_{ik}(\bmath{\theta}) A_i(t) \Big) \Big(\sum_{i=2}^N c_{ik}(\bmath{\theta}) A_i(t) \Big)^{\dagger} \bigg \rangle \, ,
    \end{split}
\end{equation}
where the notation $C_{\gamma}$ denotes the covariance matrix of the coefficients $\gamma_k(\bmath{\theta}, t)$.

\noindent The SCAO systems compensate for the zeroth order of the distortions, thus the contribution of modes higher than the tilt has to be considered. By exploiting the covariance properties of the Zernike and through straightforward algebra, it can be demonstrated that the phase variance becomes:
\begin{equation}
    \begin{split}
        \sigma_{\varphi}^2 =& \sum_{k=2}^3 \bigg \langle \Big(\sum_{i=4}^N c_{ik}(\bmath{\theta}) A_i(t) \Big) \Big(\sum_{i=4}^N c_{ik}(\bmath{\theta}) A_i(t) \Big)^{\dagger} \bigg \rangle \\
        =& \big[(2 \sqrt{3} \theta_x)^2 + (2 \sqrt{3} \theta_y)^2 \big] \langle A_4(t) A_4^{\dagger}(t)\rangle \\
        & \:\:\: + \big[(\sqrt{6} \theta_y)^2 + (\sqrt{6} \theta_x)^2 \big] \langle A_5(t) A_5^{\dagger}(t)\rangle \\
        & \:\:\: + \big[(\sqrt{6} \theta_x)^2 + (-\sqrt{6} \theta_y)^2 \big] \langle A_6(t) A_6^{\dagger}(t)\rangle + \cdots \\
        =& 6 \big(2 \langle A_4(t) A_4^{\dagger}(t)\rangle + \langle A_5(t) A_5^{\dagger}(t)\rangle \\
        &\:\:\: + \langle A_6(t) A_6^{\dagger}(t)\rangle \big) (\theta_x^2 + \theta_y^2) + \cdots \, ,
    \end{split}
\end{equation}
where we showed the results from the first order distortions. In this case the variance shows to be, at the first order, proportional to the second power of the off-axis separation (i.e. the RMS has a linear dependence).

\noindent The NGS-based MCAO configuration that we considered is able to compensate for the first order distortions. The contribution of the uncorrected modes, in this case the ones higher than the astigmatisms, leads to a phase variance that is, at the first order, proportional to the fourth power of the off-axis separation (i.e. RMS proportional to the second power):
\begin{equation}
    \begin{split}
        \sigma_{\varphi}^2 =& \sum_{k=2}^3 \bigg \langle \Big(\sum_{i=7}^N c_{ik}(\bmath{\theta}) A_i(t) \Big) \Big(\sum_{i=7}^N c_{ik}(\bmath{\theta}) A_i(t) \Big)^{\dagger} \bigg \rangle \\
        =& 18 \big(10 \langle A_8(t) A_8^{\dagger}(t) \rangle + \langle A_9(t) A_9^{\dagger}(t) \rangle \\
        & \:\:\: + \langle A_{10}(t) A_{10}^{\dagger}(t) \rangle\big) (\theta_x^2 + \theta_y^2)^2 + \cdots \, .
    \end{split}
\end{equation}

\bsp	
\label{lastpage}
\end{document}